\documentclass[pdflatex,secnum,leqno]{rims-bessatsu}
\usepackage{amssymb, amsmath}
\usepackage{eepic}
\usepackage{epic}
\usepackage{times}
\usepackage{graphicx}
\theoremstyle{definition}

\theoremstyle{remark}
\title{Integrable discretizations of the SIR model}
\author{Yuta \textsc{Tanaka}\footnote{Department of Pure and Applied Mathematics, School of Fundamental Science and Engineering, Waseda University, 3-4-1 Okubo, Shinjuku-ku, Tokyo 169-8555, Japan.}
          ~and Ken-ichi \textsc{Maruno}\footnote{Department of Applied Mathematics, Faculty of Science and Engineering, Waseda University, 3-4-1 Okubo, Shinjuku-ku, Tokyo 169-8555, Japan. \endgraf e-mail: \texttt{kmaruno@waseda.jp}}}
\AuthorHead{Yuta Tanaka and Ken-ichi Maruno}         
\VolumeNo{x}           
\YearNo{202x}           
\PagesNo{000--000}      
\communication{Received February 15, 2023. Revised September 4, 2023.}      
\begin{document}
%

\maketitle

\begin{abstract}      
Structure-preserving discretizations of the SIR model are presented by
focusing on the hodograph transformation and
the conditions for integrability for their discrete SIR models are given.
For those integrable discrete SIR models,
we derive their exact solutions as well as conserved quantities.
If we choose the parameter appropriately for one of our proposed
discrete SIR models,
it conserves the conserved quantities of the SIR model.
We also investigate an ultradiscretizable discrete SIR model.
\end{abstract}

\section{Introduction}

A simple mathematical model predicting
the behavior of epidemic outbreaks was proposed by Kermack and McKendrick in 1927\cite{SIR}.
Their mathematical epidemic model is called the Susceptible-Infected-Recovered (SIR) model.
The SIR model is composed of three differential equations for
$S$, $I$ and $R$, where they are numbers for susceptible, infected and recovered, respectively:
\begin{eqnarray}
&& \frac{dS}{dt}(t)=-\beta S(t) I(t)\,,\label{SIR1}\\
&& \frac{dI}{dt}(t)=\beta S(t) I(t) -\gamma I(t)\,,\label{SIR2}\\
&& \frac{dR}{dt}(t)=\gamma I(t)\,,\label{SIR3}
\end{eqnarray}
where $\beta$ is the infection rate and $\gamma$ is the recovery rate,
and both parameters take positive values.
Since the SIR model was proposed by Kermack and McKendrick,
this mathematical model and its extensions have been
used to analyze actual infectious diseases~\cite{Brauer,Martcheva,Hethcode}.
Most mathematical studies of the SIR model have been done
from the numerical and analytical point of view
because the exact solution of the SIR model
was unknown until recently. In 2014, Harko, Lobo and Mak
presented the exact solution to the initial value problem
for the SIR model in a parametric form\cite{Harko}.

Discretizations of the SIR model and other epidemic models have been studied
from numerical and analytical point of view in the past three decades\cite{Allen,Wacker,Moghadas}.
Among many works about discretizations of epidemic models,
Moghadas et al. \cite{Moghadas} considered a positivity-preserving Mickens-type
nonstandard discretization, i.e., one of structure-preserving discretizations,
of an epidemic model by following the idea of Mickens\cite{Mickens}.

In the studies of integrable systems, integrable discretizations, which preserve
the structure of mathematical properties such as
exact solutions and conserved quantities of integrable systems such as soliton equations,
have been actively studied and
many discrete integrable systems have been presented\cite{Discrete,Suris,Ablowitz,Joshi}.
For the SIR model, Willox et al. considered structure-preserving
discretizations of the SIR model and their ultradiscretizations\cite{Willox1,Willox2,Murata}.
Sekiguchi et al. considered an ultradiscrete SIR model with time-delay and investigated
its analytical properties\cite{Sekiguchi}.

In this paper, we present structure-preserving discretizations,
including integrable discretizations as special cases, of the SIR model which have
conserved quantities and exact solutions, by focusing on the hodograph transformation.
The key of our structure-preserving discretizations is the fact that
the SIR model is linearizable by a hodograph transformation.
In our previous papers about integrable discretizations of soliton equations such as
the Camassa-Holm equation and the short pulse equation, hodograph transformations
have played an important role in integrable discretizations, and
those integrable discretizations provided self-adaptive moving mesh schemes which
are difference schemes that produce finer meshes at locations with large deformations\cite{CH1,CH2,SP1,SP2,SP3}.

The present paper is organized as follows.
In section 2, we derive conserved quantities of the SIR model and
construct the exact solution to the initial value problem for the SIR model by using
a hodograph transformation.
In section 3, we propose three structure-preserving discretizations
of the SIR model and present the conditions for integrability,
and construct their conserved quantities and exact solutions to the initial value problem.
In section 4, we consider an ultradiscretizable SIR model and its ultradiscretization.
Section 5 is devoted to conclusions.

\section{Construction of conserved quantities and exact solutions of the SIR model by the hodograph transformation}

In this section, we construct conserved quantities and the exact solution to
the initial value problem for the SIR model
by using a hodograph transformation.

Adding the SIR model (\ref{SIR1}), (\ref{SIR2}), (\ref{SIR3}), we verify that
the total population
\begin{equation}
N=S(t)+I(t)+R(t)
\end{equation}
is conserved.

We can also find another conserved quantity in the following way.
We obtain
\begin{equation}
  \frac{d}{dt}\log S(t)=-\beta I(t) \label{logS}
\end{equation}
from equation (\ref{SIR1})
and
\begin{equation}
\frac{d}{dt}(S(t)+I(t))=-\gamma I(t) \label{SI-dif}
\end{equation}
by adding equations (\ref{SIR1}) and (\ref{SIR2}).
Then we obtain
\begin{equation}
\frac{d}{dt}\left(\beta (S(t)+I(t))-\gamma \log S(t)\right)=0\,.
\end{equation}
from (\ref{logS}) and (\ref{SI-dif}).
Thus
\begin{equation}
\beta (S(t)+I(t))-\gamma \log S(t)
\end{equation}
is conserved.

Next we construct the exact solution of the SIR model.
Note that the following construction is different from the construction
by Harko et al\cite{Harko} and is similar to the construction
by Miller\cite{Miller1,Miller2}.
We consider the hodograph transformation (reciprocal transformation)
\begin{equation}
\tau=\tau_0+\int_{t_0}^tI(\tilde{t})d\tilde{t}\label{tau-trans}
\end{equation}
and the inverse hodograph transformation
\begin{equation}
 t=t_0+\int_{\tau_0}^{\tau}\frac{1}{I(\tilde{\tau})}d\tilde{\tau}\,.
 \label{t-trans}
\end{equation}
Differentiating (\ref{tau-trans}), we obtain
\begin{equation}
\frac{d\tau}{dt}=I(t)\,.
\end{equation}
Applying the above hodograph transformation to the SIR model, we obtain
\begin{eqnarray}
&& \frac{dS}{d\tau}=-\beta S\,, \label{S}\\
&& \frac{dI}{d\tau}=\beta S  -\gamma\,, \label{I}\\
&& \frac{dR}{d\tau}=\gamma \,, \label{R}
\end{eqnarray}
which is the system of linear ordinary differential equations.
Thus we can easily find the general solution of the system of linear
differential equations (\ref{S}), (\ref{I}), (\ref{R}) as follows:
\begin{equation}
 S(\tau)=C_1e^{-\beta \tau}\,, \quad
I(\tau)=-C_1 e^{-\beta \tau}+\beta C_2 -\gamma \tau\,, \quad
R(\tau)=\gamma \tau+C_3\,.
\end{equation}
Thus, for the initial value problem of (\ref{S}), (\ref{I}), (\ref{R}) with
the initial value
\begin{equation}
 S(t_0)=S(\tau_0)=S_0\,,\quad I(t_0)=I(\tau_0)=I_0\,, \quad R(t_0)=R(\tau_0)=R_0\,,\label{initialcond}
\end{equation}
the solution is given by
\begin{eqnarray}
&&S(\tau)=S_0e^{-\beta (\tau-\tau_0)}\,, \\
&&I(\tau)=-S_0e^{-\beta (\tau-\tau_0)}+S_0+I_0-\gamma (\tau-\tau_0)\,, \\
&&R(\tau)=R_0+\gamma (\tau-\tau_0)\,.
\end{eqnarray}
Combining this solution with the hodograph transformation
(\ref{tau-trans}) and (\ref{t-trans}),
we obtain the solution to the initial value problem for
the SIR model (\ref{SIR1}), (\ref{SIR2}), (\ref{SIR3})
with the initial condition
(\ref{initialcond}) as follows:
\begin{eqnarray}
&&S(t)=S(\tau)=S_0e^{-\beta (\tau-\tau_0)}\,, \\
&&I(t)=I(\tau)=-S_0e^{-\beta (\tau-\tau_0)}+S_0+I_0-\gamma (\tau-\tau_0)\,, \\
&&R(t)=R(\tau)=R_0+\gamma (\tau-\tau_0)\,, \\
&&t=t_0+\int_{\tau_0}^{\tau}\frac{1}{I(\tilde{\tau})}d\tilde{\tau}\,.
\end{eqnarray}
Setting $u=e^{-\beta (\tau-\tau_0)}$,
the solution is written in the parametric form
\begin{eqnarray}
&&S(t)=S(\tau)=S_0u\,, \\
&&I(t)=I(\tau)=\frac{\gamma}{\beta}\log u-S_0u+S_0+I_0\,, \\
&&R(t)=R(\tau)=R_0-\frac{\gamma}{\beta}\log u\,,
\end{eqnarray}
which corresponds to the exact solution obtained by Harko et al\cite{Harko}.

\begin{figure}[htb]
  \begin{minipage}[b]{0.48\columnwidth}
  \includegraphics[width=7.5cm,pagebox=cropbox]{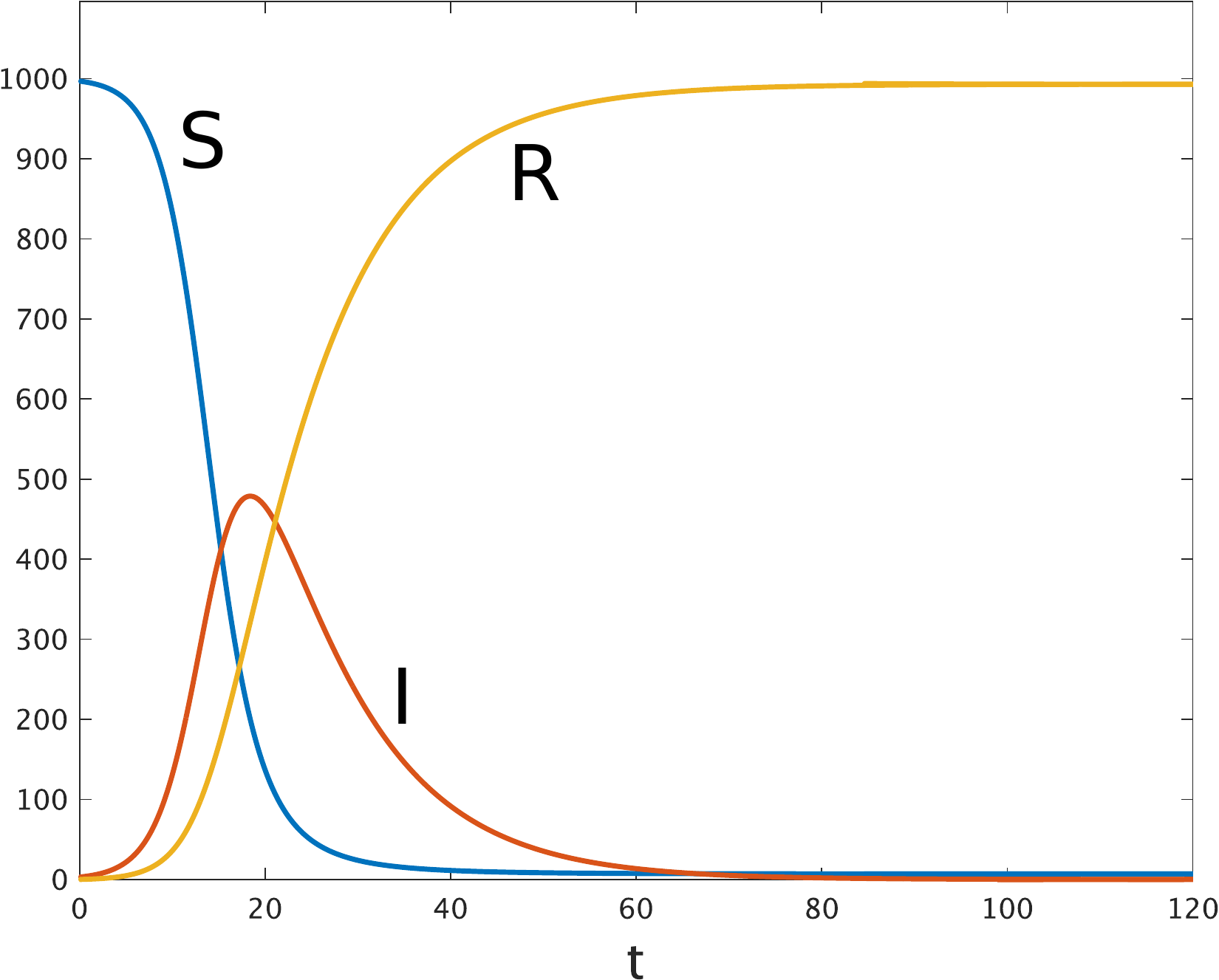}
  \end{minipage}
   \hspace{0.04\columnwidth}
  \begin{minipage}[b]{0.48\columnwidth}
  \includegraphics[width=7.5cm,pagebox=cropbox]{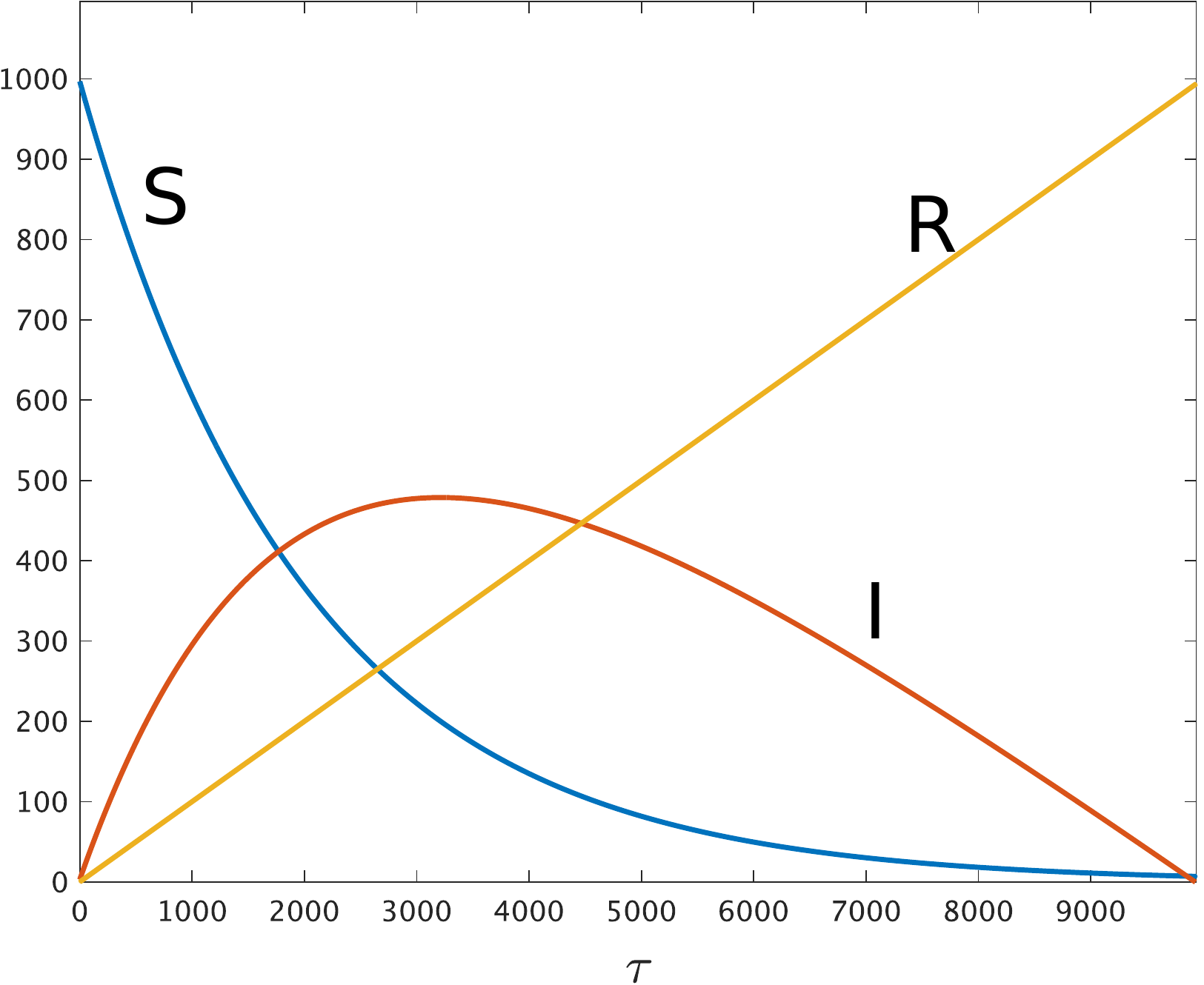}
\end{minipage}
\caption{The graphs of the exact solution to the initial value problem for the SIR model.
The parameters and initial values are $\beta=0.0005$, $\gamma=0.1$, $S(0)=997$, $I(0)=3$, $R(0)=0$.
The horizontal axis in the left panel is $t$, the horizontal axis in the right panel is $\tau$.}
\label{fig:SIRexact}
\end{figure}

In figure~\ref{fig:SIRexact}, we show the graphs of
the exact solution to the initial value problem for the SIR model.
In the right graph the horizontal axis is $\tau$ and
we note that the graph of $R$ is linear.

\section{Integrable discretizations of the SIR model}

In this section, we present structure-preserving discretizations of the SIR model
by focusing on the hodograph transformation and the conditions for integrability.
For the integrable cases of all models
we construct conserved quantities and exact solutions to the initial value problem.

\subsection{The discrete SIR-1 model}

In this subsection, we present an integrable discrete SIR model
and construct its conserved quantities and
the exact solution to the initial value problem for the discrete SIR model.

By discretizing the system of linear differential equations
(\ref{S}), (\ref{I}), (\ref{R}),
we obtain
\begin{eqnarray}
&& \frac{S_{n+1}-S_n}{\epsilon_n}=-\beta S_n\,, \label{dS}\\
&& \frac{I_{n+1}-I_n}{\epsilon_n}=\beta S_n  -\gamma\,, \label{dI}\\
&& \frac{R_{n+1}-R_n}{\epsilon_n}=\gamma\,, \label{dR}
\end{eqnarray}
where  $S_n=S(\tau_n)$, $I_n=I(\tau_n)$, $R_n=R(\tau_n)$.
Let us define $\tau_n$ and $t_n$ as
\begin{equation}
\tau_n=\tau_0+\sum_{k=0}^{n-1}\epsilon_k\label{taudef}
\end{equation}
and
\begin{equation}
t_n=t_0+\sum_{k=0}^{n-1}\delta_k\label{tdef}
\end{equation}
where $\epsilon_n$ and $\delta_n$ are lattice interval parameters which depends on $n$.
Here we note that the relations
$\delta_n=t_{n+1}-t_n$ and $\epsilon_n=\tau_{n+1}-\tau_n$ are hold.

Then we consider the discrete hodograph transformation
\begin{equation}
\tau_n=\tau_0+\sum_{k=0}^{n-1}I_k\delta_k\,,\label{dhodograph}
\end{equation}
and the inverse discrete hodograph transformation
\begin{equation}
 t_n=t_0+\sum_{k=0}^{n-1}\frac{1}{I_k}\epsilon_k\,,\label{inversedhodograph}
\end{equation}
which are discretizations of
the hodograph transformation (\ref{tau-trans}) and
its inverse hodograph transformation (\ref{t-trans}), respectively.
Then we note
\begin{equation}
t_{n+1}-t_n=\delta_n=\frac{1}{I_n}\epsilon_n\,,
\end{equation}
which leads to $\epsilon_n=\delta_nI_n$.

Substituting $\epsilon_n=\delta_nI_n$ into the system of linear
difference equations
(\ref{dS}), (\ref{dI}), (\ref{dR}),
we obtain
\begin{eqnarray}
&& \frac{S_{n+1}-S_n}{\delta_n}\frac{1}{I_n}=-\beta S_n\,, \\
&& \frac{I_{n+1}-I_n}{\delta_n}\frac{1}{I_n}=\beta S_n  -\gamma\,, \\
&& \frac{R_{n+1}-R_n}{\delta_n}\frac{1}{I_n}=\gamma\,,
\end{eqnarray}
which leads to a discretization of the SIR model
\begin{eqnarray}
&& \frac{S_{n+1}-S_n}{\delta_n}=-\beta S_nI_n\,, \label{dSIR1}\\
&& \frac{I_{n+1}-I_n}{\delta_n}=\beta S_n I_n -\gamma I_n\,, \label{dSIR2}\\
&& \frac{R_{n+1}-R_n}{\delta_n}=\gamma I_n\label{dSIR3}\,,\\
&& t_n=t_0+\sum_{k=0}^{n-1}\delta_k=t_0+\sum_{k=0}^{n-1}\frac{1}{I_k}\epsilon_k\,,\nonumber
\end{eqnarray}
where
$S_n=S(t_n)=S(\tau_n)$, $I_n=I(t_n)=I(\tau_n)$,
$R_n=R(t_n)=R(\tau_n)$.
Hereafter we refer (\ref{dSIR1}), (\ref{dSIR2}), (\ref{dSIR3}) as
the discrete SIR-1 (dSIR1) model.
Note that the set of $(t_n,S_n)$,
$(t_n,I_n)$, $(t_n,R_n)$ provides the approximate solution of the SIR model.

Next we consider conserved quantities of the dSIR1 model.
From (\ref{dSIR1}), (\ref{dSIR2}), (\ref{dSIR3}), we obtain
\begin{equation}
S_{n+1}+I_{n+1}+R_{n+1}=S_n+I_n+R_n\,.
\end{equation}
Thus $S_n+I_n+R_n$ is a conserved quantity of the
dSIR1 model (\ref{dSIR1}), (\ref{dSIR2}), (\ref{dSIR3}).

From equation (\ref{dS}) we obtain
\begin{equation}
 \log S_{n+1}=\log (1-\beta \epsilon_n)+\log S_n
\end{equation}
which leads to
\begin{equation}
\log S_{n+1}-\log S_n=\log (1-\beta \epsilon_n)\,.\label{logS:dS1}
\end{equation}
Adding (\ref{dS}) and (\ref{dI}), we obtain
\begin{equation}
S_{n+1}+I_{n+1}-S_n-I_n=-\gamma \epsilon_n\,.\label{SI:dS1}
\end{equation}
From (\ref{logS:dS1}) and (\ref{SI:dS1}), we obtain
\begin{equation}
(S_{n+1}+I_{n+1}-S_n-I_n)\log(1-\beta \epsilon_n)
+ \gamma \epsilon_n(\log S_{n+1}-\log S_n)
=0\,.
\end{equation}
Since
\begin{equation}
(S_{n+1}+I_{n+1}-S_n-I_n) \log(1-\beta \epsilon_n)
+ \gamma \epsilon_n(\log S_{n+1}-\log S_n)\label{invariant:lineardS1}
\end{equation}
is zero for any $n$,
this is an invariant for (\ref{dS}), (\ref{dI}), (\ref{dR}).
Substituting $\epsilon_n=\delta_nI_n$ into (\ref{invariant:lineardS1}),
we obtain
\begin{equation}
(S_{n+1}+I_{n+1}-S_n-I_n) \log(1-\beta \delta_n I_n)
+ \gamma \delta_n I_n(\log S_{n+1}-\log S_n)=0\label{Invariant:dSIR1}
\end{equation}
which is an invariant for the dSIR1 model
(\ref{dSIR1}), (\ref{dSIR2}), (\ref{dSIR3}).

If we set $\epsilon_n=\epsilon$, where $\epsilon$ is a constant,
we have
\begin{equation}
 (S_{n+1}+I_{n+1})\log(1-\beta \epsilon)
+ \gamma \epsilon \log S_{n+1}
=\log(1-\beta \epsilon)(S_n+I_n)
+ \gamma \epsilon \log S_n\label{Invariant2:dSIR1}
\end{equation}
which indicates that
\begin{equation}
(S_{n}+I_{n})\log(1-\beta \epsilon)+\gamma \epsilon \log S_n \label{CQ:lineardS1}
\end{equation}
is a conserved quantity of (\ref{dS}), (\ref{dI}), (\ref{dR}).
Substituting $\epsilon=\delta_n I_n$ into (\ref{CQ:lineardS1}),
we obtain
\begin{equation}
(S_{n}+I_{n})\log(1-\beta \delta_n  I_n)+\gamma \delta_n I_n \log S_n
\label{dSIR-CQ}
\end{equation}
which is a conserved quantity of dSIR1 model (\ref{dSIR1}), (\ref{dSIR2}), (\ref{dSIR3}).
This means that the dSIR1 model is integrable when $\epsilon_n$ is a constant.
Note that the dSIR1 model is the forward Euler scheme of the SIR model
when $\delta_n$ is a constant, but there is no second conserved quantity in this case, i.e.,
the forward Euler scheme of the SIR model is nonintegrable.
Although the forward Euler scheme of the SIR model is nonintegrable,
this scheme has the invariant (\ref{Invariant:dSIR1}) which is reduced to the conserved quantity in
the integrable case. The existence of the invariant (\ref{Invariant:dSIR1})
indicates near-integrability of the dSIR1 model when $\epsilon_n$ is not constant.

Next we verify directly that (\ref{dSIR-CQ}) is a conserved quantity
of the dSIR1 model when
$\epsilon_n$ is a constant.
From (\ref{dSIR1}), we have
\begin{equation}
S_{n+1}=(1-\beta \delta_n I_n )S_n\,,
\end{equation}
and by taking the logarithm of both sides of this equation we obtain
\begin{equation}
 \log S_{n+1}-\log S_n= \log (1-\beta \delta_n I_n )\,.\label{dSIR-C1}
\end{equation}
Adding (\ref{dSIR1}) and (\ref{dSIR2}), we obtain
\begin{equation}
S_{n+1}+I_{n+1}-(S_n+I_n)=-\gamma \delta_n I_n\,.\label{dSIR-C2}
\end{equation}
Combining (\ref{dSIR-C1}) and (\ref{dSIR-C2}), we obtain
\begin{equation}
(S_{n+1}+I_{n+1}-(S_n+I_n)) \log (1-\beta \delta_n  I_n)
+\gamma \delta_n I_n (\log S_{n+1}-\log S_n)=0
\end{equation}
which does not depend on $n$.
Setting $\epsilon_n=\delta_nI_n=\epsilon$,
we obtain
\begin{equation}
(S_{n+1}+I_{n+1}) \log (1-\beta \epsilon )
+\gamma \epsilon \log S_{n+1}=(S_{n}+I_{n})\log (1-\beta \epsilon )
+\gamma \epsilon \log S_{n}\,.
\end{equation}
Thus
\begin{equation}
 (S_{n}+I_{n}) \log (1-\beta \epsilon )
+\gamma \epsilon \log S_{n}
\end{equation}
is a conserved quantity of the dSIR1 model.
By taking the limit $\delta_n\to 0$, we obtain
\begin{equation}
\beta (S(t)+I(t))-\gamma \log S(t)
\end{equation}
which is a conserved quantity of the SIR model.

Let us consider the exact solution to the initial value problem for
the dSIR1 model.
For the initial value
$S(t_0)=S(\tau_0)=S_0$, $I(t_0)=I(\tau_0)=I_0$, $R(t_0)=R(\tau_0)=R_0$,
the exact solution of the system of
linear difference equations (\ref{dS}), (\ref{dI}), (\ref{dR}) is given by
\begin{align}
S_n&=S_0\prod_{k=0}^{n-1}(1-\beta \epsilon_{k})\,,\\
I_n&=I_0+\beta \sum_{k=0}^{n-1}\epsilon_k
S_{k}-\gamma \sum_{k=0}^{n-1}\epsilon_k=I_0+\beta S_0\sum_{k=0}^{n-1}\epsilon_k
\prod_{l=0}^{k-1}(1-\beta \epsilon_{l})
-\gamma \sum_{k=0}^{n-1}\epsilon_k\,,\\
R_n&=R_0+\gamma \sum_{k=0}^{n-1}\epsilon_k\,.
\end{align}
Substituting $\epsilon_n=\delta_n I_n$ into this solution,
the exact solution to the initial value problem for the dSIR1 model
(\ref{dSIR1}), (\ref{dSIR2}), (\ref{dSIR3}) is obtained:
\begin{align}
S_n&=S_0\prod_{k=0}^{n-1}(1-\beta \delta_{k}I_{k})\,,\\
I_n
&=I_0+\beta \sum_{k=0}^{n-1} \delta_kI_k S_{k}-\gamma \sum_{k=0}^{n-1}\delta_kI_k\\
&=I_0+\beta S_0\sum_{k=0}^{n-1}\delta_kI_k
\prod_{l=0}^{k-1}(1-\beta \delta_{l}I_{l})
-\gamma \sum_{k=0}^{n-1}\delta_kI_k\,,\notag\\
R_n&=R_0+\gamma \sum_{k=0}^{n-1} \delta_k I_k\,,\\
t_n&=t_0+\sum_{k=0}^{n-1}\delta_k\,.
\end{align}
By using (\ref{dSIR2}), $I_n$ can be also written as
\begin{equation}
I_n=I_0\prod_{k=0}^{n-1} (1-\gamma \delta_k+\beta \delta_k S_k)\,.
\end{equation}
Since $S_n$ and $I_n$ always take positive values in the SIR model,
we find two inequalities
\begin{eqnarray}
I_k<\frac{1}{\beta \delta_k}\,, \quad
S_k>\frac{\gamma}{\beta}-\frac{1}{\beta \delta_k}
\end{eqnarray}
which must be satisfied when we use the dSIR1 model as a numerical scheme.

If we set $\epsilon_n=\epsilon$, i.e., integrable case,
the above exact solution
leads to the following simple form:
\begin{eqnarray}
&&S_n=S_0(1-\beta \epsilon)^{n}\,,\\
&&I_n=I_0+S_0(1-(1-\beta \epsilon)^{n})-\gamma \epsilon n
=S_0+I_0-S_0(1-\beta \epsilon)^{n}-\gamma \epsilon n\,, \\
&&R_n=R_0+\gamma \epsilon n\,,\\
&&t_n
=t_0+\sum_{k=0}^{n-1}\frac{1}{I_k}\epsilon\,.
\end{eqnarray}
Note that $S_n$ is written in the form of a power function which
includes the infection rate $\beta$,
the lattice parameter $\epsilon$ and
the initial value $S_0$, and $I_n$ is a linear combination of a power function and a linear function.
This drastic simplification is due to integrability.
Since $S_n$ and $I_n$ always take positive values,
we find two inequalities
\begin{eqnarray}
&&\beta \epsilon<1\,, \quad
S_0(1-\beta\epsilon)^n+\gamma \epsilon n<S_0+I_0
\end{eqnarray}
which must be satisfied when we use the dSIR1 model with $\epsilon_k=\epsilon$
as a numerical scheme.

If we set $\delta_n=\delta$, where $\delta$ is a constant,
the above exact solution leads to
\begin{align}
S_n&=S_0\prod_{k=0}^{n-1}(1-\beta \delta I_{k})\,,\\
I_n&=I_0+\beta \delta \sum_{k=0}^{n-1} I_k S_k
-\gamma \delta \sum_{k=0}^{n-1}I_k
=I_0+\beta S_0 \delta
\sum_{k=0}^{n-1}I_k \prod_{l=0}^{k-1}(1-\beta \delta I_{l})
-\gamma \delta \sum_{k=0}^{n-1}I_k\\
&=I_0\prod_{k=0}^{n-1} (1-\gamma \delta+\beta \delta S_k)\,, \notag\\
R_n&=R_0+\gamma \delta \sum_{k=0}^{n-1}I_k\,,\\
t_n&=t_0+n\delta\,.
\end{align}

\begin{figure}[htb]
  \begin{minipage}[b]{0.48\columnwidth}
  \includegraphics[width=7.5cm,pagebox=cropbox]{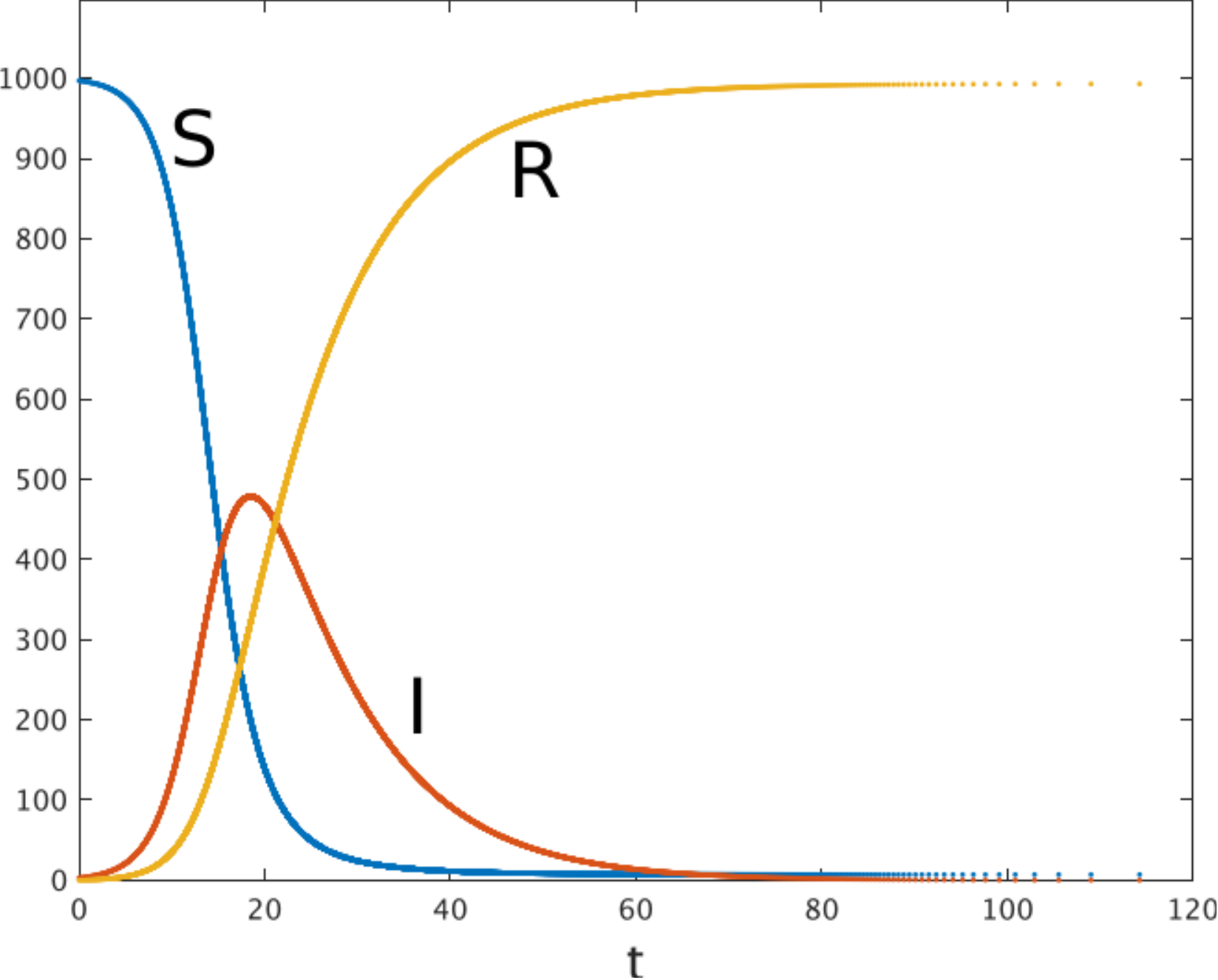}
  \end{minipage}
  \hspace{0.04\columnwidth}
  \begin{minipage}[b]{0.48\columnwidth}
  \includegraphics[width=7.5cm,pagebox=cropbox]{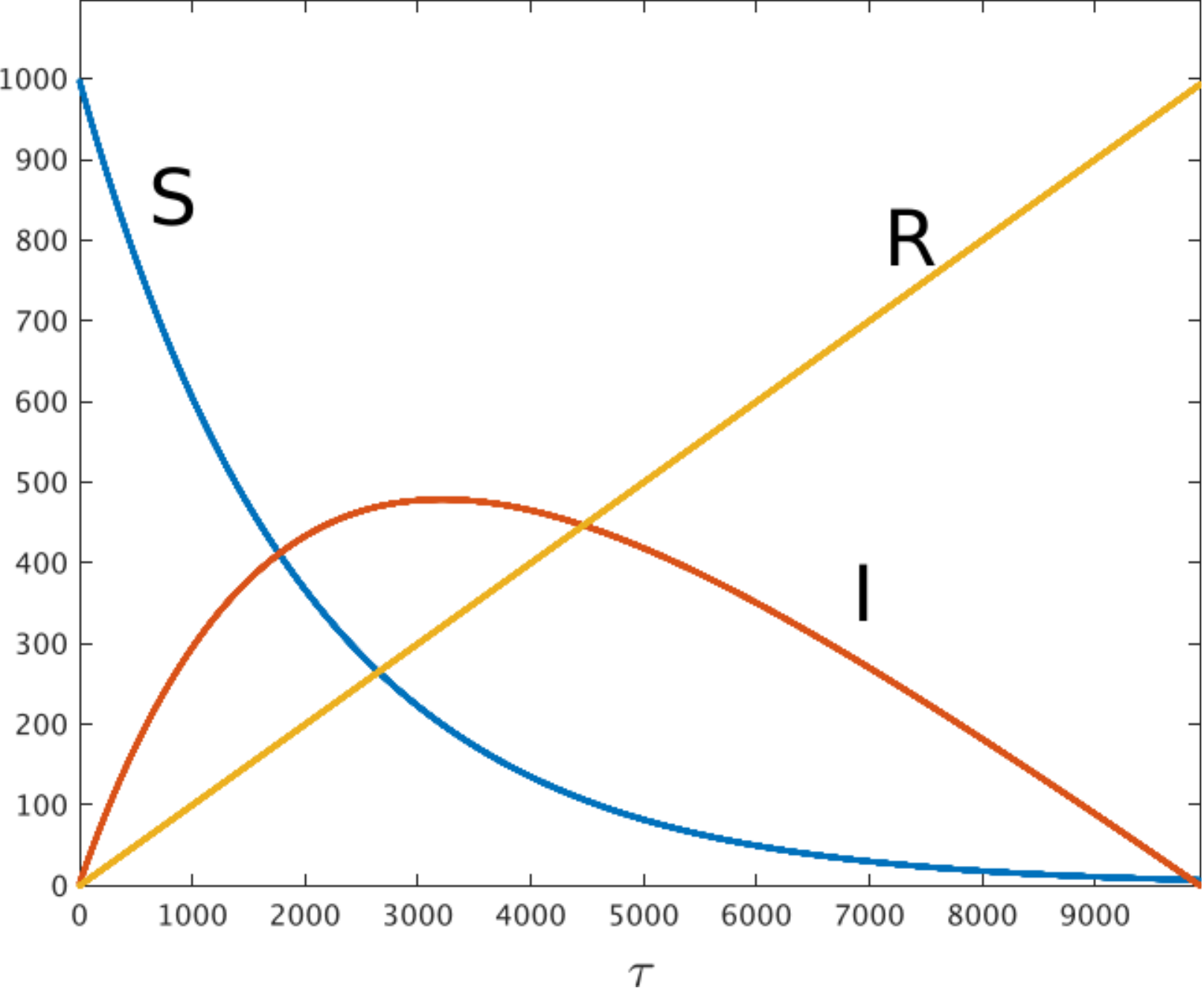}
\end{minipage}
\caption{The graphs of an exact solution to the initial value problem for the dSIR1 model.
The parameters and initial values are
$\beta=0.0005, \gamma=0.1, S(0)=997, I(0)=3, R(0)=0$, $\epsilon_k=0.5$.
The horizontal axis in the left panel is $t$, the horizontal axis in the right panel is $\tau$.}
\label{fig:dSIR1exact}
\end{figure}

In figure~\ref{fig:dSIR1exact}, we show the graphs of
the exact solution to the initial value problem for the dSIR1 model in the case of
$\epsilon_n=\epsilon$.
As you can see from the area around the right of the left panel in figure~\ref{fig:dSIR1exact},
the dSIR1 model generates finer meshes where $I_n$ is large.
This is the same as the characteristics of self-adaptive moving mesh schemes.

\subsection{The discrete SIR-2 model}

In this subsection, we present another integrable discrete SIR model and
construct its conserved quantities and
the exact solution to the initial value problem of the discrete SIR model.

Here we consider the following discretization of
the system of linear differential equations
(\ref{S}), (\ref{I}), (\ref{R}):
\begin{eqnarray}
&& \frac{S_{n+1}-S_{n}}{\epsilon_n}=-\beta S_{n+1}\,, \label{dSback}\\
&& \frac{I_{n+1}-I_n}{\epsilon_n}=\beta S_{n+1}  -\gamma \label{dIback}\,,\\
&& \frac{R_{n+1}-R_n}{\epsilon_n}=\gamma\,, \label{dRback}
\end{eqnarray}
where  $S_n=S(\tau_n)$, $I_n=I(\tau_n)$, $R_n=R(\tau_n)$.
As in the case of the dSIR1 model, let us define $\tau_n$ and $t_n$ as
(\ref{taudef}) and (\ref{tdef}).
Then we consider the discrete hodograph transformation
(\ref{dhodograph}) and the inverse discrete
hodograph transformation (\ref{inversedhodograph}).

Substituting $\epsilon_n=\delta_nI_n$ into the system of linear
difference equations (\ref{dSback}), (\ref{dIback}), (\ref{dRback}),
we obtain
\begin{eqnarray}
&& \frac{S_{n+1}-S_n}{\delta_n}\frac{1}{I_n}=-\beta S_{n+1}\,, \\
&& \frac{I_{n+1}-I_n}{\delta_n}\frac{1}{I_n}=\beta S_{n+1}  -\gamma \,,\\
&& \frac{R_{n+1}-R_n}{\delta_n}\frac{1}{I_n}=\gamma\,,
\end{eqnarray}
which leads to a discretization of the SIR model
\begin{eqnarray}
&& \frac{S_{n+1}-S_n}{\delta_n}=-\beta S_{n+1}I_n\,, \label{dSIRback1}\\
&& \frac{I_{n+1}-I_n}{\delta_n}=\beta S_{n+1} I_n -\gamma I_{n} \,,\label{dSIRback2}\\
&& \frac{R_{n+1}-R_n}{\delta_n}=\gamma I_{n}\,,\label{dSIRback3}\\
&& t_n=t_0+\sum_{k=0}^{n-1}\delta_k=t_0+\sum_{k=0}^{n-1}\frac{1}{I_k}\epsilon_k\,,\nonumber
\end{eqnarray}
where
$S_n=S(t_n)=S(\tau_n)$, $I_n=I(t_n)=I(\tau_n)$,
$R_n=R(t_n)=R(\tau_n)$.
Hereafter we refer (\ref{dSIRback1}), (\ref{dSIRback2}), (\ref{dSIRback3}) as
the discrete SIR-2 (dSIR2) model.
Note that the dSIR2 model
(\ref{dSIRback1}), (\ref{dSIRback2}), (\ref{dSIRback3}) is rewritten as
\begin{eqnarray}
&& S_{n+1}=\frac{S_n}{1+\beta \delta_n I_n}\,,\label{dSIRback2-1}\\
&& I_{n+1}=(1+\beta \delta_n S_{n+1}-\gamma \delta_n)I_n\,,
\label{dSIRback2-2}\\
&& R_{n+1}=R_n+\gamma \delta_n I_{n}\,,\label{dSIRback2-3}
\end{eqnarray}
and (\ref{dSIRback2-2}) is rewritten as
\begin{equation}
 I_{n+1}=\frac{1-\gamma \delta_n +\beta \delta_n(1-\gamma \delta_n)I_n+\beta \delta_n S_n}
{1+\beta \delta_n I_n}I_n\,.
\end{equation}
Note that the set of $(t_n,S_n)$,
$(t_n,I_n)$, $(t_n,R_n)$ provides the approximate solution of the SIR model.

Next we consider conserved quantities.
As in the case of the dSIR1 model,
$S_n+I_n+R_n$ is a conserved quantity of the
dSIR2 model (\ref{dSIRback1}), (\ref{dSIRback2}), (\ref{dSIRback3}).

From equation (\ref{dSback}) we obtain
\begin{equation}
 S_{n+1}=\frac{S_n}{1+\beta \epsilon_n}
\end{equation}
which leads to
\begin{equation}
 \log S_{n+1}=\log S_n-\log (1+\beta \epsilon_n)\,.\label{logS:dS2}
\end{equation}
Adding (\ref{dSback}) and (\ref{dIback}), we obtain
\begin{equation}
S_{n+1}+I_{n+1}-S_n-I_n=-\gamma \epsilon_n\,.\label{}\label{SI:dS2}
\end{equation}
From (\ref{logS:dS2}) and (\ref{SI:dS2}), we obtain
\begin{equation}
(S_{n+1}+I_{n+1}-S_n-I_n)\log(1+\beta \epsilon_n)
- \gamma \epsilon_n(\log S_{n+1}-\log S_n)
=0\,.
\end{equation}
Since
\begin{equation}
 (S_{n+1}+I_{n+1}-S_n-I_n)\log(1+\beta \epsilon_n)
- \gamma \epsilon_n(\log S_{n+1}-\log S_n)
\end{equation}
is zero for any $n$,
this is an invariant for (\ref{dSback}), (\ref{dIback}), (\ref{dRback}).
Substituting $\epsilon_n=\delta_nI_n$ into this invariant,
we obtan
\begin{equation}
(S_{n+1}+I_{n+1}-S_n-I_n)\log(1+\beta \delta_n I_n)
- \gamma \delta_n I_n(\log S_{n+1}-\log S_n)=0\label{Invariant:dSIR2}
\end{equation}
which is an invariant for the dSIR2 model (\ref{dSIRback1}),
(\ref{dSIRback2}), (\ref{dSIRback3}).

If we set $\epsilon_n=\epsilon$, where $\epsilon$ is a constant,
we have
\begin{equation}
(S_{n+1}+I_{n+1})\log(1+\beta \epsilon)
-\gamma \epsilon \log S_{n+1}
=(S_n+I_n)\log(1+\beta \epsilon)
- \gamma \epsilon \log S_n
\end{equation}
which indicates that
\begin{equation}
(S_{n}+I_{n})\log(1+\beta \epsilon)-\gamma \epsilon \log S_n
\end{equation}
is a conserved quantity of (\ref{dSback}), (\ref{dIback}), (\ref{dRback}).
Substituting $\epsilon=\delta_n I_n$ into this,
we obtain
\begin{equation}
(S_{n}+I_{n})\log(1+\beta \delta_n I_n )-\gamma \delta_n  I_n\log S_n
\label{dSIRback-CQ}
\end{equation}
which is a conserved quantity of the dSIR2 model (\ref{dSIRback1}), (\ref{dSIRback2}), (\ref{dSIRback3}). This means that the dSIR2 model is
integrable when $\epsilon_n$ is a constant.
Other cases including $\delta_n=\delta$, where $\delta$ is a constant, are nonintegrable because
there is no second conserved quantity.
The dSIR2 model has the invariant (\ref{Invariant:dSIR2}) which is reduced to the conserved quantity in
the integrable case. The existence of the invariant (\ref{Invariant:dSIR2})
indicates near-integrability of the dSIR2 model.

Next we verify directly that (\ref{dSIRback-CQ}) is a conserved quantity of the dSIR2 model
when $\epsilon_n$ is a constant.
From (\ref{dSIRback1}), we have
\begin{equation}
 S_{n+1}=\frac{S_n}{1+\beta \delta_n I_n}
\end{equation}
and by taking the logarithm of both sides of this equation we obtain
\begin{equation}
 \log S_{n+1}-\log S_n= -\log (1+\beta \delta_n I_n)\,.\label{dSIR2-C1}
\end{equation}
Adding (\ref{dSIRback1}) and (\ref{dSIRback2}), we obtain
\begin{equation}
S_{n+1}+I_{n+1}-(S_n+I_n)=-\gamma \delta_n I_n\,.\label{dSIR2-C2}
\end{equation}
Combining (\ref{dSIR2-C1}) and (\ref{dSIR2-C2}), we obtain
\begin{equation}
(S_{n+1}+I_{n+1}-(S_n+I_n)) \log (1-\beta \delta_n  I_n)
+\gamma \delta_n I_n (\log S_{n+1}-\log S_n)=0
\end{equation}
which does not depend on $n$.
Setting $\epsilon_n=\delta_nI_n=\epsilon$,
we obtain
\begin{equation}
 (S_{n+1}+I_{n+1})\log (1+\beta \epsilon )
-\gamma \epsilon \log S_{n+1}=(S_{n}+I_{n})\log (1+\beta \epsilon )
-\gamma \epsilon \log S_{n}
\end{equation}
Thus
\begin{equation}
 (S_{n}+I_{n})\log (1+\beta \epsilon )
-\gamma \epsilon \log S_{n}
\end{equation}
is a conserved quantity.
By taking the limit $\delta_n\to 0$, we obtain
\begin{equation}
\beta (S(t)+I(t))-\gamma \log S(t)
\end{equation}
which is a conserved quantity of the SIR model.

Let us consider the exact solution to the initial value problem for the dSIR2 model.
For the initial value
$S(t_0)=S(\tau_0)=S_0$, $I(t_0)=I(\tau_0)=I_0$, $R(t_0)=R(\tau_0)=R_0$,
the exact solution of the system of linear difference equations
(\ref{dSback}), (\ref{dIback}), (\ref{dRback}) is given
by
\begin{eqnarray}
&&S_n=S_0\prod_{k=0}^{n-1}\frac{1}{1+\beta \epsilon_{k}}\,,\\
&&I_n=I_0+\beta \sum_{k=0}^{n-1}\epsilon_k
S_{k+1}
-\gamma \sum_{k=0}^{n-1}\epsilon_k\\
&&\quad =I_0+\beta S_0\sum_{k=0}^{n-1}\epsilon_k
\prod_{l=0}^{k}\frac{1}{1+\beta \epsilon_{l}}
-\gamma \sum_{k=0}^{n-1}\epsilon_k\,,\nonumber\\
&&R_n=R_0+\gamma \sum_{k=0}^{n-1}\epsilon_k\,.
\end{eqnarray}
Substituting $\epsilon_n=\delta_n I_n$ into this solution,
the exact solution to the initial value problem for the
dSIR2 model
(\ref{dSIRback1}), (\ref{dSIRback2}), (\ref{dSIRback3})
is obtained:
\begin{eqnarray}
&&S_n=S_0\prod_{k=0}^{n-1}\frac{1}{1+\beta \delta_{k}I_{k}}\,,\\
&&I_n=I_0+\beta \sum_{k=0}^{n-1} \delta_kI_k S_{k+1}
-\gamma \sum_{k=0}^{n-1}\delta_kI_k\\
&&\quad =I_0+\beta S_0\sum_{k=0}^{n-1}\delta_kI_k
\prod_{l=0}^{k}\frac{1}{1+\beta \delta_{l}I_{l}}
-\gamma \sum_{k=0}^{n-1}\delta_kI_k\,,\nonumber\\
&&R_n=R_0+\gamma \sum_{k=0}^{n-1} \delta_k I_k\,,\\
&&t_n=t_0+\sum_{k=0}^{n-1}\delta_k\,.
\end{eqnarray}
By using (\ref{dSIRback2}), $I_n$ can be also written as
\begin{equation}
I_n=I_0\prod_{k=0}^{n-1} (1-\gamma \delta_k+\beta \delta_k S_{k+1})\,.
\end{equation}
In the dSIR2 model, $S_n$ is always positive.
Since $S_n$ and $I_n$ always take positive values in the SIR model,
we find an inequality
\begin{eqnarray}
&&S_{k+1}>\frac{\gamma}{\beta}-\frac{1}{\beta \delta_k}
\end{eqnarray}
which must be satisfied when we use the dSIR2 model as a numerical scheme.

If we set $\epsilon_n=\epsilon$, i.e., integrable case,
the above exact solution takes the following simple form:
\begin{eqnarray}
&&S_n=\frac{S_0}{(1+\beta \epsilon)^{n}}\,,\\
&&I_n=S_0+I_0-S_0(1+\beta \epsilon)^{-n}-\gamma \epsilon n\,, \\
&&R_n=R_0+\gamma \epsilon n\,,\\
&&t_n=t_0+\sum_{k=0}^{n-1}\frac{1}{I_k}\epsilon\,.
\end{eqnarray}
Note that $S_n$ is written in the form of a power function which includes the infection rate $\beta$,
the lattice parameter $\epsilon$ and
the initial value $S_0$, and $I_n$ is a linear combination of a power function and a linear function.
This drastic simplification is due to integrability.
Since $S_n$ and $I_n$ always take positive values in the SIR model,
we find an inequality
\begin{eqnarray}
&&S_0(1+\beta\epsilon)^{-n}+\gamma \epsilon n<S_0+I_0
\end{eqnarray}
which must be satisfied when we use the dSIR2 model with $\epsilon_n=\epsilon$
as a numerical scheme.

If we set $\delta_n=\delta$, where $\delta$ is a constant, the above exact solution leads to
\begin{eqnarray}
&&S_n=S_0\prod_{k=0}^{n-1}\frac{1}{1+\beta \delta I_{k}}\,,\\
&&I_n=I_0+\beta \delta \sum_{k=0}^{n-1} I_k S_{k+1}
-\gamma \delta \sum_{k=0}^{n-1}I_k\\
&&\quad =I_0+\beta S_0 \delta
\sum_{k=0}^{n-1}I_k \prod_{l=0}^{k}\frac{1}{1+\beta \delta I_{l}}
-\gamma \delta \sum_{k=0}^{n-1}I_k\,,\nonumber\\
&&\quad =I_0\prod_{k=0}^{n-1} (1-\gamma \delta+\beta \delta S_{k+1})\,,\nonumber \\
&&R_n=R_0+\gamma \delta \sum_{k=0}^{n-1}I_k\,,\\
&&t_n=t_0+n\delta\,.
\end{eqnarray}

\begin{figure}[htb]
  \begin{minipage}[b]{0.48\columnwidth}
  \includegraphics[width=7.5cm,pagebox=cropbox]{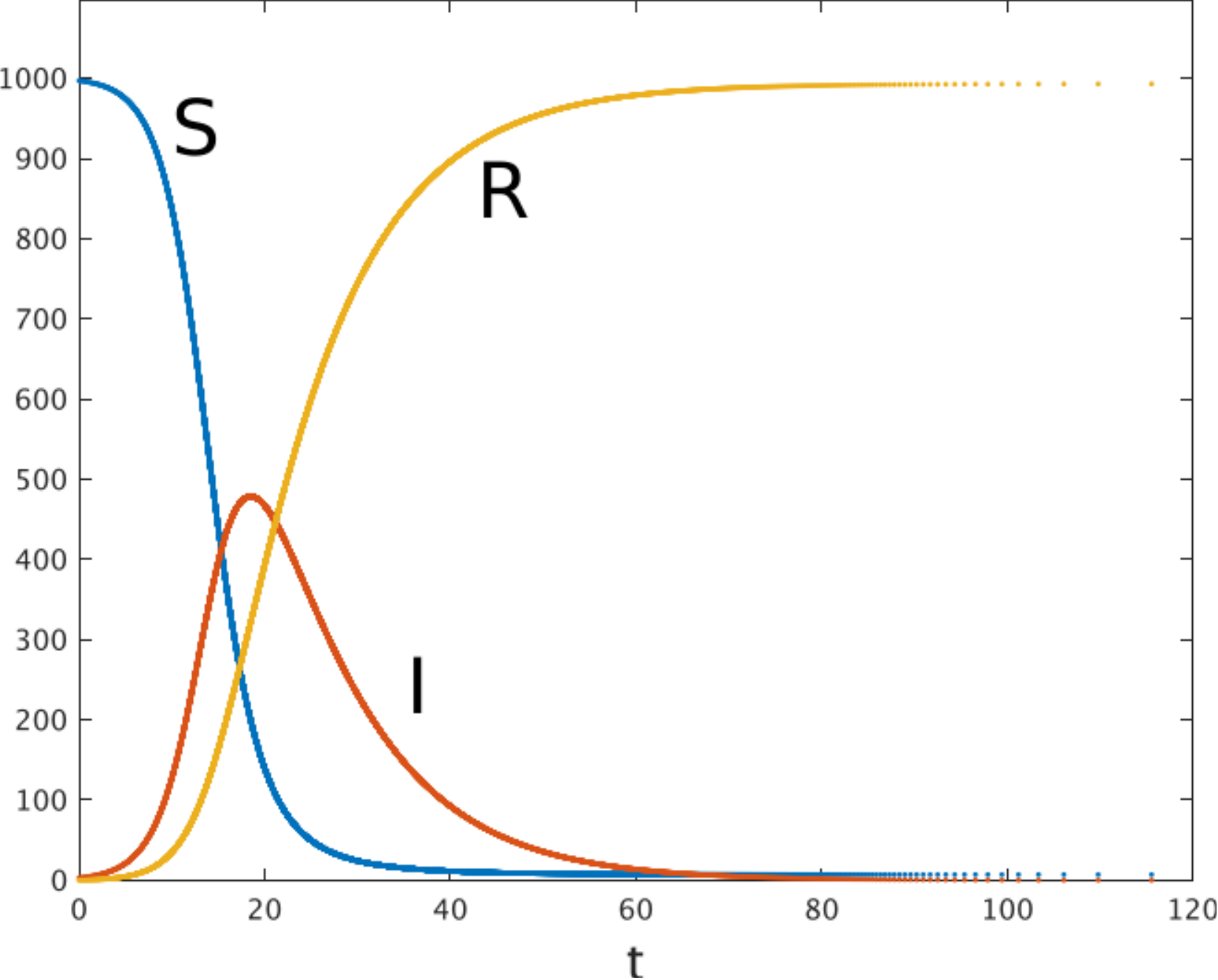}
  \end{minipage}
  \hspace{0.04\columnwidth}
  \begin{minipage}[b]{0.48\columnwidth}
  \includegraphics[width=7.5cm,pagebox=cropbox]{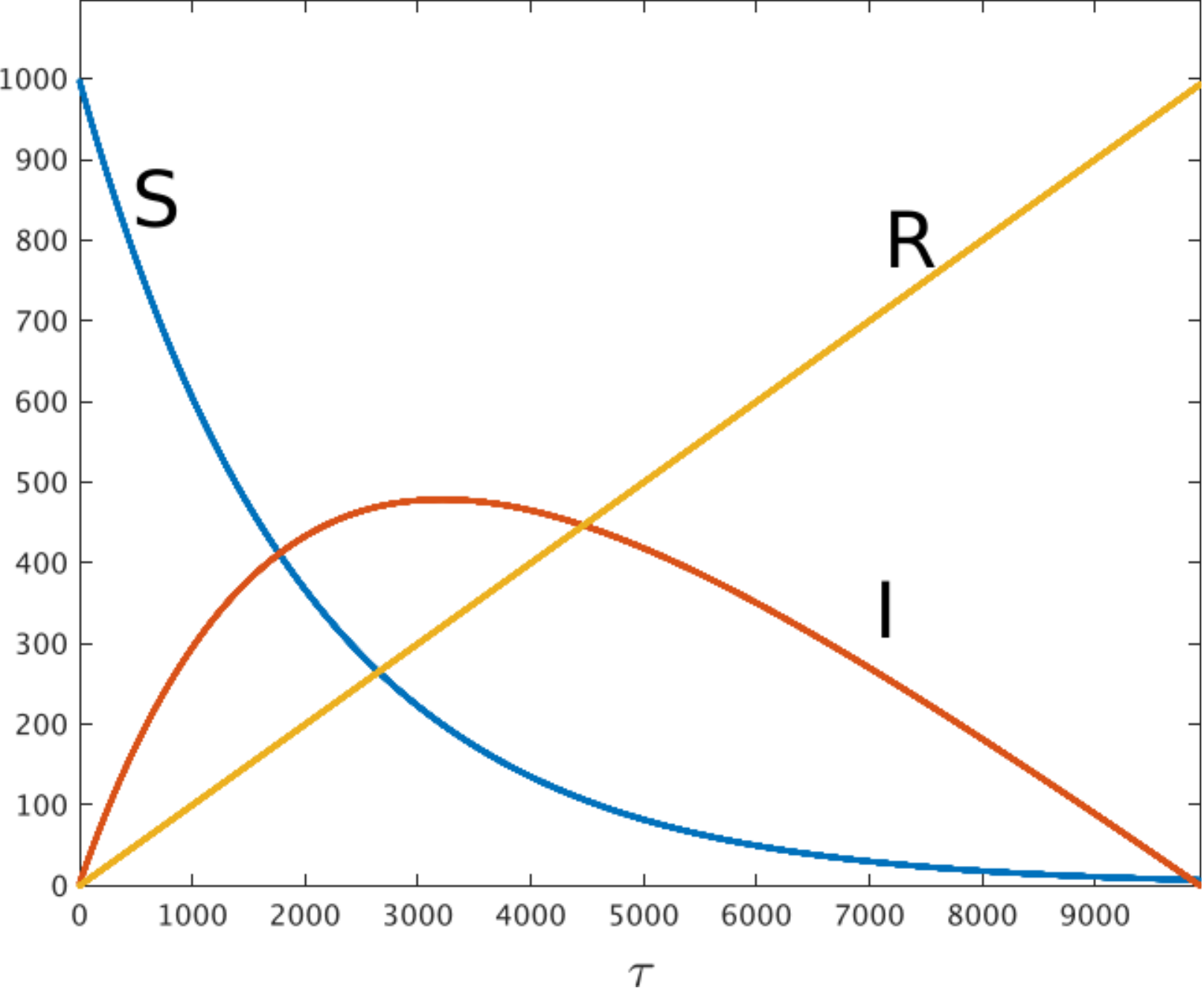}
\end{minipage}
\caption{The graphs of an exact solution to the initial value problem for the dSIR2 model.
The parameters and initial values are
$\beta=0.0005, \gamma=0.1, S(0)=997, I(0)=3, R(0)=0$, $\epsilon_k=0.5$.
The horizontal axis in the left panel is $t$, the horizontal axis in the right panel is $\tau$.}
\label{fig:dSIR2exact}
\end{figure}

In figure~\ref{fig:dSIR2exact}, we show the graphs of
the exact solution to the initial value problem for the dSIR2 model in the case of
$\epsilon_n=\epsilon$.
As you can see from the area around the right of the left panel in figure~\ref{fig:dSIR2exact},
the dSIR2 model generates finer meshes where $I_n$ is large.
This is the same as the characteristics of self-adaptive moving mesh schemes.

\subsection{The generalized discrete SIR model}

In this subsection, we generalize previous two discrete SIR models and
construct its conserved quantities and
the exact solution to the initial value problem for the discrete SIR model.

We discretize the system of linear differential equations
(\ref{S}), (\ref{I}), (\ref{R}) in the following form:
\begin{eqnarray}
&& \frac{S_{n+1}-S_{n}}{\epsilon_n}=-\beta (pS_{n+1}+(1-p)S_n)\,, \label{gdS}\\
&& \frac{I_{n+1}-I_n}{\epsilon_n}=\beta (pS_{n+1}+(1-p)S_n)  -\gamma\,, \label{gdI}\\
&& \frac{R_{n+1}-R_n}{\epsilon_n}=\gamma \,,\label{gdR}
\end{eqnarray}
where  $S_n=S(\tau_n)$, $I_n=I(\tau_n)$, $R_n=R(\tau_n)$,
and the parameter $p$ is a real number between 0 and 1.
This can be written as
\begin{eqnarray}
&& \frac{S_{n+1}-S_{n}}{\epsilon_n}=-\beta M_p S_n\,, \\
&& \frac{I_{n+1}-I_n}{\epsilon_n}=\beta M_p S_n  -\gamma\,,\\
&& \frac{R_{n+1}-R_n}{\epsilon_n}=\gamma \,,
\end{eqnarray}
by using the weighted averaging operator $M_p$ defined by
\begin{equation}
 M_pf_n=pf_{n+1}+(1-p)f_n\,.
\end{equation}
Let us define $\tau_n$ and $t_n$ as
(\ref{taudef}) and (\ref{tdef}).
Then we consider the discrete hodograph transformation
(\ref{dhodograph}) and
the inverse discrete hodograph transformation (\ref{inversedhodograph}).

Substituting $\epsilon_n=\delta_nI_n$ into the system of linear
difference equations (\ref{gdS}), (\ref{gdI}), (\ref{gdR}),
we obtain
\begin{eqnarray}
&& \frac{S_{n+1}-S_n}{\delta_n}\frac{1}{I_n}=-\beta (pS_{n+1}+(1-p)S_n)\,,\\
&& \frac{I_{n+1}-I_n}{\delta_n}\frac{1}{I_n}=\beta (pS_{n+1}+(1-p)S_n)  -\gamma\,,\\
&& \frac{R_{n+1}-R_n}{\delta_n}\frac{1}{I_n}=\gamma\,,
\end{eqnarray}
which leads to a discretization of the SIR model
\begin{eqnarray}
&& \frac{S_{n+1}-S_n}{\delta_n}=-\beta (pS_{n+1}+(1-p)S_n)I_n\,, \label{gdSIR1}\\
&& \frac{I_{n+1}-I_n}{\delta_n}=\beta (pS_{n+1}+(1-p)S_n) I_n -\gamma I_{n}\,,\label{gdSIR2}\\
&& \frac{R_{n+1}-R_n}{\delta_n}=\gamma I_{n}\,,\label{gdSIR3}
\end{eqnarray}
where
$S_n=S(t_n)=S(\tau_n)$, $I_n=I(t_n)=I(\tau_n)$,
$R_n=R(t_n)=R(\tau_n)$.
Hereafter we refer (\ref{gdSIR1}), (\ref{gdSIR2}), (\ref{gdSIR3}) as
the generalized discrete SIR (gdSIR) model.
Note that the gdSIR model
(\ref{gdSIR1}), (\ref{gdSIR3}), (\ref{gdSIR3}) is rewritten as
\begin{eqnarray}
  && S_{n+1}=S_n\frac{1-(1-p)\beta \delta_n I_n}{1+p\beta \delta_n I_n}\,,\label{gdSIR2-1}\\
  && I_{n+1}=\left(1+\beta \delta_n (pS_{n+1}+(1-p)S_n)-\gamma \delta_n\right)I_n\,,
  \label{gdSIR2-2}\\
  && R_{n+1}=R_n+\gamma \delta_n I_{n+1}\label{dSIR2-3:2}\,.
\end{eqnarray}
Note that the set of $(t_n,S_n)$,
$(t_n,I_n)$, $(t_n,R_n)$ provides the approximate solution of the SIR model.
The gdSIR model (\ref{gdSIR1}), (\ref{gdSIR2}), (\ref{gdSIR3}) becomes
the dSIR1 model in the case of $p=0$ and
the dSIR2 model in the case of $p=1$.

Next we consider conserved quantities.
We can easily see that $S_n+I_n+R_n$ is a conserved quantity of the
gdSIR model (\ref{gdSIR1}), (\ref{gdSIR2}), (\ref{gdSIR3}).

From equation (\ref{gdS}) we obtain
\begin{equation}
 S_{n+1}=\frac{1-(1-p)\beta \epsilon_n}{1+p\beta \epsilon_n}S_n
\end{equation}
which leads to
\begin{equation}
 \log S_{n+1}=\log S_n+\log\frac{1-(1-p)\beta \epsilon_n}{1+p\beta \epsilon_n}\,.\label{logS:gdS}
\end{equation}
Adding (\ref{gdS}) and (\ref{gdI}), we obtain
\begin{equation}
S_{n+1}+I_{n+1}-S_n-I_n=-\gamma \epsilon_n\label{SI:gdS}
\end{equation}
From (\ref{logS:gdS}) and (\ref{SI:gdS}), we obtain
\begin{equation}
(S_{n+1}+I_{n+1}-(S_n+I_n))\log\frac{1-(1-p)\beta \epsilon_n }{1+p\beta \epsilon_n}
  +\gamma \epsilon_n(\log S_{n+1}-\log S_n)=0\,.
\end{equation}
Since
\begin{equation}
(S_{n+1}+I_{n+1}-(S_n+I_n))\log\frac{1-(1-p)\beta \epsilon_n}{1+p\beta \epsilon_n}
  +\gamma \epsilon_n (\log S_{n+1}-\log S_n)
\end{equation}
is zero for any $n$,
this is an invariant for (\ref{gdS}), (\ref{gdI}), (\ref{gdR}).
Substituting $\epsilon_n=\delta_nI_n$ into this invariant,
we obtan
\begin{equation}
 (S_{n+1}+I_{n+1}-S_n-I_n)\log\frac{1-(1-p)\beta \delta_n I_n}{1+p\beta \delta_n I_n}
+ \gamma \delta_n I_n(\log S_{n+1}-\log S_n)=0
\end{equation}
which is an invariant for the gdSIR model (\ref{gdSIR1}),
(\ref{gdSIR2}), (\ref{gdSIR3}).

If we set $\epsilon_n=\epsilon$, where $\epsilon$ is a constant,
we have
\begin{eqnarray}
&&(S_{n+1}+I_{n+1})\log\frac{1-(1-p)\beta \epsilon}{1+p\beta \epsilon}
  + \gamma \epsilon(\log S_{n+1})\\
&&\qquad =(S_n+I_n)\log\frac{1-(1-p)\beta \epsilon}{1+p\beta \epsilon}
  + \gamma \epsilon(\log S_n)\nonumber
\end{eqnarray}
which indicates that
\begin{equation}
  (S_{n}+I_{n})\log\frac{1-(1-p)\beta \epsilon}{1+p\beta \epsilon}
  +\gamma \epsilon \log S_n
\end{equation}
is a conserved quantity of (\ref{gdS}), (\ref{gdI}), (\ref{gdR}).
Substituting $\epsilon=\delta_n I_n$ into this,
we obtain
\begin{equation}
  (S_{n}+I_{n})\log\frac{1-(1-p)\beta \delta_n I_n}{1+p\beta \delta_n I_n}
  +\gamma \delta_n  I_n\log S_n
\label{gdSIR-CQ}
\end{equation}
which is a conserved quantity of the gdSIR model
(\ref{gdSIR1}), (\ref{gdSIR2}), (\ref{gdSIR3}).
This means that the gdSIR model
is integrable when $\epsilon_n$ is a constant.
Other cases including $\delta_n=\delta$, where $\delta$ is a constant, are nonintegrable.

Next we verify directly that (\ref{gdSIR-CQ}) is a conserved quantity
of the gdSIR model when $\epsilon_n$ is a constant.
From (\ref{gdSIR1}), we have
\begin{equation}
S_{n+1}=S_n\frac{1-(1-p)\beta \delta_n I_n}{1+p\beta \delta_n I_n}
\end{equation}
and by taking the logarithm of both sides of this equation we obtain
\begin{equation}
 \log S_{n+1}-\log S_n= \log \frac{1-(1-p)\beta \delta_n I_n}{1+p\beta \delta_n I_n}
 \label{gdSIR-C1}
\end{equation}
Adding (\ref{gdSIR1}) and (\ref{gdSIR2}), we obtain
\begin{equation}
S_{n+1}+I_{n+1}-(S_n+I_n)=-\gamma \delta_n I_n\,.\label{gdSIR-C2}
\end{equation}
Combining (\ref{gdSIR-C1}) and (\ref{gdSIR-C2}), we obtain
\begin{align}
&(S_{n+1}+I_{n+1}-(S_n+I_n))\log\frac{1-(1-p)\beta \delta_n I_n}{1+p\beta \delta_n I_n}\\
&\qquad +\gamma \delta_n  I_n (\log S_{n+1}-\log S_n)=0\notag
\end{align}
which does not depend on $n$.
Setting $\epsilon_n=\delta_nI_n=\epsilon$,
we obtain
\begin{eqnarray}
  &&(S_{n+1}+I_{n+1})\log \frac{1-(1-p)\beta \epsilon}{1+p\beta \epsilon}
  +\gamma \epsilon \log S_{n+1}\\
  &&\qquad =(S_{n}+I_{n})\log \frac{1-(1-p)\beta \epsilon}{1+p\beta \epsilon}
  +\gamma \epsilon \log S_{n}\,.\nonumber
\end{eqnarray}
Thus
\begin{equation}
  (S_{n}+I_{n})\log \frac{1-(1-p)\beta \epsilon}{1+p\beta \epsilon}
  +\gamma \epsilon \log S_{n}
\end{equation}
is a conserved quantity.
By taking the limit $\delta_n\to 0$, we obtain
\begin{equation}
\beta (S(t)+I(t))-\gamma \log S(t)
\end{equation}
which is a conserved quantity of the SIR model.

Let us consider the solution to the initial value problem for the
gdSIR model (\ref{gdSIR1}), (\ref{gdSIR2}), (\ref{gdSIR3}).
For the initial value
$S(t_0)=S(\tau_0)=S_0$, $I(t_0)=I(\tau_0)=I_0$, $R(t_0)=R(\tau_0)=R_0$,
the solution of (\ref{gdS}), (\ref{gdI}), (\ref{gdR}) is given
by
\begin{align}
S_n&=S_0\prod_{k=0}^{n-1}\frac{1-(1-p)\beta \epsilon_k}{1+p\beta \epsilon_{k}}\,,\\
I_n&=I_0+\beta \sum_{k=0}^{n-1}\epsilon_k (pS_{k+1}+(1-p)S_k)
  -\gamma \sum_{k=0}^{n-1}\epsilon_k\\
&=I_0-\gamma \sum_{k=0}^{n-1}\epsilon_k\notag\\
&\quad+\beta S_0\sum_{k=0}^{n-1}\epsilon_k
  \left(p\prod_{l=0}^{k}\frac{1-(1-p)\beta \epsilon_l}{1+p\beta \epsilon_{l}}
  +(1-p)\prod_{l=0}^{k-1}\frac{1-(1-p)\beta \epsilon_l}{1+p\beta \epsilon_{l}}
  \right)
  \notag\\
&=I_0  -\gamma \sum_{k=0}^{n-1}\epsilon_k
+\beta S_0\sum_{k=0}^{n-1}\frac{\epsilon_k}{1+p\beta \epsilon_k}
\prod_{l=0}^{k-1}\frac{1-(1-p)\beta \epsilon_l}{1+p\beta \epsilon_{l}}
\,,\notag\\
R_n&=R_0+\gamma \sum_{k=0}^{n-1}\epsilon_k\,.
\end{align}
Substituting $\epsilon_n=\delta_n I_n$ into this solution,
the solution to the initial value problem for the
gdSIR model (\ref{gdSIR1}), (\ref{gdSIR2}), (\ref{gdSIR3})
is obtained:
\begin{align}
S_n&=S_0\prod_{k=0}^{n-1}\frac{1-(1-p)\beta \delta_kI_k}{1+p\beta \delta_{k}I_{k}}\,,\\
I_n&=I_0+\beta \sum_{k=0}^{n-1} \delta_kI_k (pS_{k+1}+(1-p)S_k)
  -\gamma \sum_{k=0}^{n-1}\delta_kI_k\\
&=I_0+
  \beta S_0\sum_{k=0}^{n-1}\frac{\delta_kI_k}{1+p\beta \delta_kI_k}
  \prod_{l=0}^{k-1}\frac{1-(1-p)\beta \delta_l I_l}{1+p\beta \delta_{l}I_{l}}
  -\gamma \sum_{k=0}^{n-1}\delta_kI_k\,,\notag\\
R_n&=R_0+\gamma \sum_{k=0}^{n-1} \delta_k I_k\,,\\
t_n&=t_0+\sum_{k=0}^{n-1}\delta_k\,.
\end{align}
By using (\ref{gdSIR2}), $I_n$ can be also written as
\begin{equation}
I_n=I_0\prod_{k=0}^{n-1} (1-\gamma \delta_k+\beta \delta_k (pS_{k+1}+(1-p)S_k))\,.
\end{equation}
Since $S_n$ and $I_n$ always take positive values in the SIR model,
we find two inequalities
\begin{equation}
I_k<\frac{1}{(1-p)\beta \delta_k} \quad \mbox{\rm for} \quad p\neq 1\,,\quad
pS_{k+1}+(1-p)S_k>\frac{\gamma}{\beta}-\frac{1}{\beta \delta_k}
\end{equation}
which must be satisfied when we use the gdSIR model as a numerical scheme.

If we set $\epsilon_n=\epsilon$, i.e., integrable case,
the above exact solution takes the following simple form:
\begin{align}
S_n&=S_0\left(\frac{1-(1-p)\beta \epsilon}{1+p\beta \epsilon}\right)^n\,,\label{gdSIR:Sexact}\\
I_n&=I_0  -\gamma \epsilon n\\
&\quad +\beta \epsilon S_0
  \left(p
  \frac{1-(1-p)\beta \epsilon}{1+p\beta \epsilon}
  \frac{1-\left(\frac{1-(1-p)\beta \epsilon}{1+p\beta \epsilon}\right)^n}
  {1-\frac{1-(1-p)\beta \epsilon}{1+p\beta \epsilon}}
  +(1-p)\frac{1-\left(\frac{1-(1-p)\beta \epsilon}{1+p\beta \epsilon}\right)^n}
  {1-\frac{1-(1-p)\beta \epsilon}{1+p\beta \epsilon}}
  \right)\notag\\
&=S_0+I_0-\gamma \epsilon n-S_0\left(\frac{1-(1-p)\beta \epsilon}{1+p\beta \epsilon}\right)^n
\,,\notag\\
R_n&=R_0+\gamma \epsilon n\,,\\
t_n&=t_0+\sum_{k=0}^{n-1}\frac{1}{I_k}\epsilon\,.
\end{align}
Note that $S_n$ is written in the form of a power function which includes the infection rate $\beta$,
the lattice parameter $\epsilon$ and
the initial value $S_0$, and $I_n$ is a linear combination of a power function and a linear function.
This drastic simplification is due to integrability.
Since $S_n$ and $I_n$ always take positive values in the SIR model,
we find two inequalities
\begin{eqnarray}
&&(1-p)\beta \epsilon <1 \quad \mbox{\rm for} \quad p\neq 1\,,\\
&&  S_0\left(\frac{1-(1-p)\beta \epsilon}{1+p\beta \epsilon}\right)^n+\gamma \epsilon n <S_0+I_0
\end{eqnarray}
which must be satisfied when we use the gdSIR model with $\epsilon_n=\epsilon$
as a numerical scheme.

If we set $\delta_n=\delta$, the above solution leads to
\begin{align}
 S_n&=S_0\prod_{k=0}^{n-1}\frac{1-(1-p)\beta \delta I_k}{1+p\beta \delta I_{k}}\,,\\
 I_n&=I_0+\beta \sum_{k=0}^{n-1} \delta I_k (pS_{k+1}+(1-p)S_k)
-\gamma \delta \sum_{k=0}^{n-1}I_k\\
 &=I_0+
  \beta S_0\sum_{k=0}^{n-1}\frac{\delta I_k}{1+p\beta \delta I_k}
  \prod_{l=0}^{k-1}\frac{1-(1-p)\beta \delta I_l}{1+p\beta \delta I_{l}}
  -\gamma \delta \sum_{k=0}^{n-1} I_k\notag\\
 &=I_0\prod_{k=0}^{n-1} (1-\gamma \delta+\beta \delta S_{k+1})\,, \notag\\
 R_n&=R_0+\gamma \delta \sum_{k=0}^{n-1}I_k\,.
\end{align}

\begin{figure}[htb]
  \centering
  \includegraphics[width=10.0cm,pagebox=cropbox]{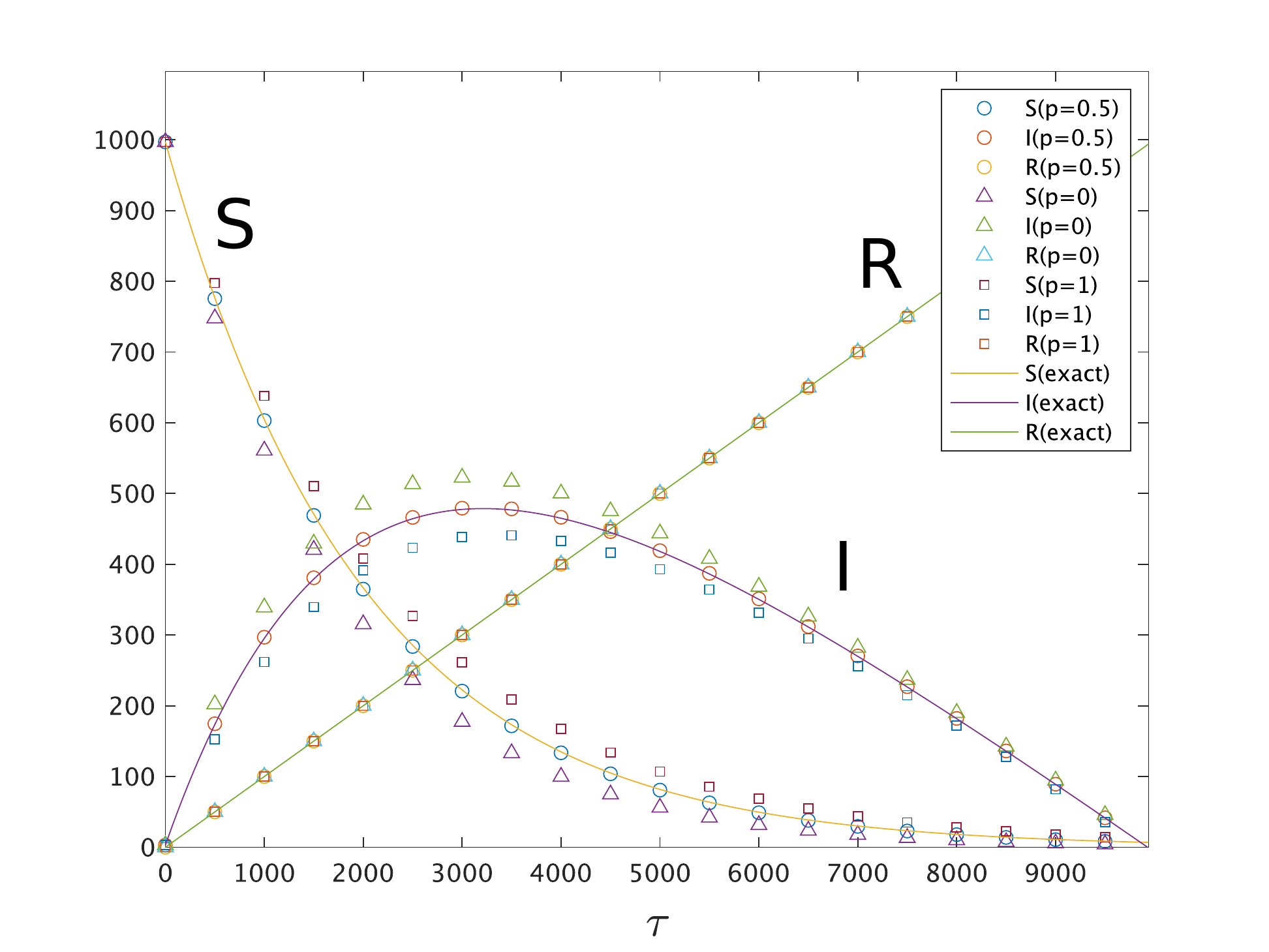}
\caption{The graph of an exact solution to the initial value problem for the
gdSIR model with
$p=0$
(the dSIR1 model),
$p=0.5$, $p=1$ (the dSIR2 model). The parameters and initial values are
$\beta=0.0005, \gamma=0.1, S(0)=997, I(0)=3, R(0)=0$, $\epsilon_k=500$.
The horizontal axis is $\tau$.}
\label{fig:gdSIRexact}
\end{figure}

In figure~\ref{fig:gdSIRexact}, we show the graphs of
the exact solution to the initial value problems for the dSIR1 model,
the dSIR2 model and the gdSIR model with $p=0.5$ in the case of
$\epsilon_n=\epsilon$, where the horizontal axis is $\tau$.
Among these three cases, the gdSIR model with $p=0.5$
gives the exact solution which is very close to the exact solution of the SIR model.

To find the best value of $p$ for which
the second conserved quantities of the gdSIR model and
the SIR model coincide, the relation
\begin{equation}
 \beta (S+I)-\gamma \log S
=  -\frac{1}{\epsilon}(S+I)\log \frac{1-(1-p)\beta \epsilon}{1+p\beta \epsilon}
    -\gamma  \log S\,.
\end{equation}
must be satisfied for any solutions.
This leads to
\begin{equation}
  \frac{1-(1-p)\beta \epsilon}{1+p\beta \epsilon}=e^{-\beta \epsilon},
\end{equation}
and solving this equation, we obtain
\begin{equation}
p=\frac{\beta \epsilon -1+e^{-\beta \epsilon}}
{\beta \epsilon(1-e^{-\beta \epsilon})}
=\frac{(\beta \epsilon -1)e^{\beta \epsilon}+1}
{\beta \epsilon(e^{\beta \epsilon}-1)}
\,. \label{pformula}
\end{equation}
Thus by using this formula to determine the value $p$, the second conserved quantity of
the gdSIR model coincides with the one of the SIR model.
In the case of figure~\ref{fig:gdSIRexact},
the best value of $p$ is $0.520812\cdots$\,.
If we substitute the formula (\ref{pformula}) into the exact solution (\ref{gdSIR:Sexact}),
we
find
\begin{equation}
S_n=S_0 \left(\frac{1-(1-p)\beta \epsilon}{1+p\beta \epsilon}\right)^n
=S_0e^{-\beta \epsilon n}=S_0e^{-\beta (\tau-\tau_0)}\,,
\end{equation}
where $\tau=\tau_0+\epsilon n$.
This means that $S_n$ in the gdSIR model coincides with $S(\tau)$ in the SIR model.
Thus $S_n$, $I_n$, $R_n$ in the gdSIR model coincide with $S(\tau)$, $I(\tau)$,
$R(\tau)$ in the SIR model respectively if we choose the best value of $p$.
However, the discrete time variable $t_n$ in the gdSIR model
does not coincide with the time variable $t$ in the SIR model.
Thus the solution set $(t_n,S_n)$, $(t_n,I_n)$,
$(t_n,S_n)$ of the gdSIR model is different from the solution set
$(t,S(t))$, $(t,I(t))$, $(t, R(t))$ of the SIR model.
To obtain numerical solutions that are close to exact solutions,
it is necessary to choose $\epsilon$ as small as possible.

\begin{figure}[htb]
  \centering
  \includegraphics[width=9.0cm,pagebox=cropbox]{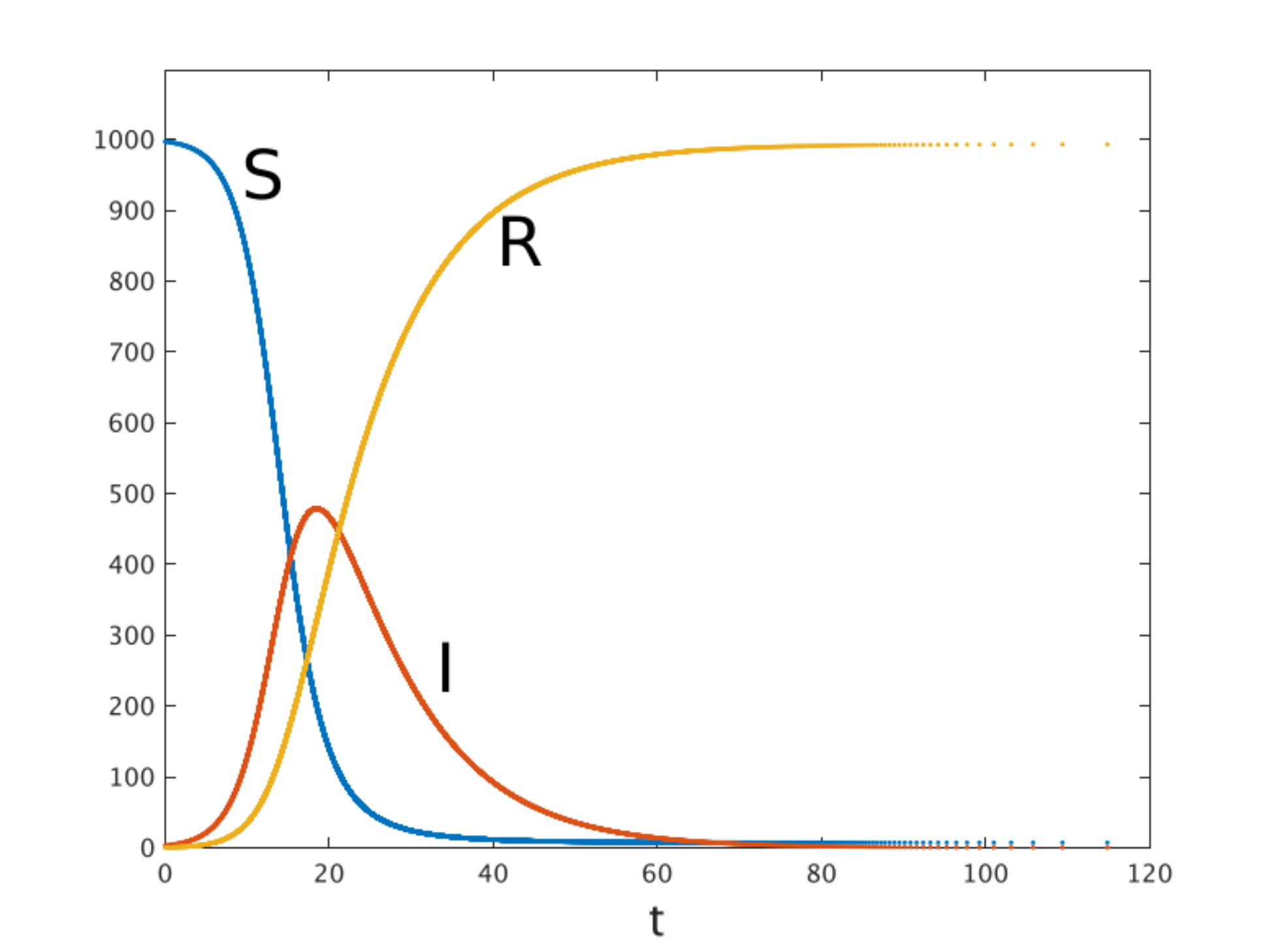}
\caption{The graph of a numerical solution to the initial value problem for the
gdSIR model with
the best value $p=0.50021\cdots$\,. The parameters and initial values are
$\beta=0.0005, \gamma=0.1, S(0)=997, I(0)=3, R(0)=0$, $\epsilon_k=0.5$.}
\label{fig:gdSIRnumerics}
\end{figure}

In figure~\ref{fig:gdSIRnumerics}, we show the graphs of
a numerical computation by the gdSIR model with the best value of $p$.
In this case, i.e.,
$\beta=0.0005$, $\epsilon_k=\epsilon=0.5$,
the best value is $p=0.50021\cdots$ which is very close to $0.5$.
This is because
\begin{equation}
  \lim_{\epsilon\to 0}\frac{(\beta \epsilon -1)e^{\beta \epsilon}+1}
  {\beta \epsilon(e^{1\beta \epsilon}-1)}=\frac{1}{2}\,.
\end{equation}
Thus we can choose $p=0.5$ for numerical computations
if we choose small enough $\epsilon$.

\subsection{The nonautonomous generalized discrete SIR model}

In this subsection, we consider the nonautonomous gdSIR model.

We discretize the system of linear differential equations
(\ref{S}), (\ref{I}), (\ref{R}) in the following form:
\begin{eqnarray}
&& \frac{S_{n+1}-S_{n}}{\epsilon_n}=-\beta (p_nS_{n+1}+(1-p_n)S_n)\,, \label{ngdS}\\
&& \frac{I_{n+1}-I_n}{\epsilon_n}=\beta (p_nS_{n+1}+(1-p_n)S_n)  -\gamma\,, \label{ngdI}\\
&& \frac{R_{n+1}-R_n}{\epsilon_n}=\gamma \,,\label{ngdR}
\end{eqnarray}
where  $S_n=S(\tau_n)$, $I_n=I(\tau_n)$, $R_n=R(\tau_n)$,
and the parameter $p_n$ are real numbers between 0 and 1 depending on $n$.
Let us define $\tau_n$ and $t_n$ as
(\ref{taudef}) and (\ref{tdef}).
Then we consider the discrete hodograph transformation
(\ref{dhodograph}) and the inverse discrete hodograph transformation (\ref{inversedhodograph}).

Substituting $\epsilon_n=\delta_nI_n$ into the system of linear
difference equations (\ref{ngdS}), (\ref{ngdI}), (\ref{ngdR}),
we obtain the nonautonomous gdSIR model
\begin{eqnarray}
&& \frac{S_{n+1}-S_n}{\delta_n}=-\beta (p_nS_{n+1}+(1-p_n)S_n)I_n\,, \label{ngdSIR1}\\
&& \frac{I_{n+1}-I_n}{\delta_n}=\beta (p_nS_{n+1}+(1-p_n)S_n) I_n -\gamma I_{n}\,,\label{ngdSIR2}\\
&& \frac{R_{n+1}-R_n}{\delta_n}=\gamma I_{n}\,,\label{ngdSIR3}
\end{eqnarray}
where
$S_n=S(t_n)=S(\tau_n)$, $I_n=I(t_n)=I(\tau_n)$,
$R_n=R(t_n)=R(\tau_n)$.
Note that the nonautonomous gdSIR model
(\ref{ngdSIR1}), (\ref{ngdSIR2}), (\ref{ngdSIR3}) is rewritten as
\begin{eqnarray}
  && S_{n+1}=S_n\frac{1-(1-p_n)\beta \delta_n I_n}{1+p_n\beta \delta_n I_n}\,,\label{ngdSIR2-1}\\
  && I_{n+1}=\left(1+\beta \delta_n (p_nS_{n+1}+(1-p_n)S_n)-\gamma \delta_n\right)I_n\,,
  \label{ngdSIR2-2}\\
  && R_{n+1}=R_n+\gamma \delta_n I_{n}\label{ndSIR2-3:2}\,.
\end{eqnarray}
Note that the set of $(t_n,S_n)$,
$(t_n,I_n)$, $(t_n,R_n)$ provides the approximate solution of the SIR model.

Next we consider conserved quantities.
We can easily see that $S_n+I_n+R_n$ is a conserved quantity of the
nonautonomous gdSIR model (\ref{ngdSIR1}), (\ref{ngdSIR2}), (\ref{ngdSIR3}).

From equation (\ref{ngdS}) we obtain
\begin{equation}
 S_{n+1}=\frac{1-(1-p_n)\beta \epsilon_n}{1+p_n\beta \epsilon_n}S_n
\end{equation}
which leads to
\begin{equation}
 \log S_{n+1}-\log S_n=\log\frac{1-(1-p_n)\beta \epsilon_n}{1+p_n\beta \epsilon_n}\,.\label{logS:ngdS}
\end{equation}
Adding (\ref{ngdS}) and (\ref{ngdI}), we obtain
\begin{equation}
S_{n+1}+I_{n+1}-S_n-I_n=-\gamma \epsilon_n\label{SI:ngdS}
\end{equation}
From (\ref{logS:ngdS}) and (\ref{SI:ngdS}), we obtain
\begin{eqnarray}
&&\beta((S_{n+1}+I_{n+1})-(S_n+I_n))-\gamma (\log S_{n+1}-\log S_n)\label{condition:ngdSIR}\\
&&\qquad =-\beta \gamma \epsilon_n-
\gamma \log\frac{1-(1-p_n)\beta \delta_n I_n}{1+p_n\beta \delta_n I_n}
  \,.\nonumber
\end{eqnarray}
If we require
\begin{equation}
\log\frac{1-(1-p_n)\beta \delta_n I_n}{1+p_n\beta \delta_n I_n}=-\beta \epsilon_n
\label{condeq:ngdSIR}
\end{equation}
to (\ref{condition:ngdSIR}),
then we obtain
\begin{equation}
\beta(S_{n+1}+I_{n+1})-\gamma \log S_{n+1}=\beta(S_n+I_n)-\gamma \log S_n
\,,
\end{equation}
which indicates that
\begin{equation}
\beta(S_n+I_n)-\gamma \log S_n \label{ngdSIR-CQ}
\end{equation}
is a conserved quantity of the
nonautonomous gdSIR model (\ref{ngdSIR1}), (\ref{ngdSIR2}), (\ref{ngdSIR3}),
but it is also a conserved quantity of the continuous SIR model.
Solving (\ref{condeq:ngdSIR}), we obtain
\begin{equation}
  p_n=\frac{\beta \epsilon_n -1+e^{-\beta \epsilon_n}}
  {\beta \epsilon_n(1-e^{-\beta \epsilon_n})}
  =\frac{(\beta \epsilon_n -1)e^{\beta \epsilon_n}+1}
  {\beta \epsilon_n(e^{\beta \epsilon_n}-1)}\,,\label{conditionformula1:ngdSIR}
\end{equation}
and substituting $\epsilon_n=\delta_nI_n$ into (\ref{conditionformula1:ngdSIR}),
this formula is written as
\begin{equation}
  p_n=\frac{\beta \delta_nI_n -1+e^{-\beta \delta_nI_n}}
  {\beta \delta_nI_n(1-e^{-\beta \delta_nI_n})}
  =\frac{(\beta \delta_nI_n -1)e^{\beta \delta_nI_n}+1}
  {\beta \delta_nI_n(e^{\beta \delta_nI_n}-1)}\,.\label{conditionformula2:ngdSIR}
\end{equation}
This means that the nonautonomous gdSIR model
is integrable when $p_n$ is given by (\ref{conditionformula2:ngdSIR}).

Next we verify directly that (\ref{ngdSIR-CQ}) is a conserved quantity
of the gdSIR model when $p_n$ is given by (\ref{conditionformula2:ngdSIR}).
From (\ref{ngdSIR1}), we have
\begin{equation}
S_{n+1}=S_n\frac{1-(1-p_n)\beta \delta_n I_n}{1+p_n\beta \delta_n I_n}
\end{equation}
and by taking the logarithm of both sides of this equation we obtain
\begin{equation}
 \log S_{n+1}-\log S_n= \log \frac{1-(1-p_n)\beta \delta_n I_n}{1+p_n\beta \delta_n I_n}
 \label{ngdSIR-C1}
\end{equation}
Adding (\ref{ngdSIR1}) and (\ref{ngdSIR2}), we obtain
\begin{equation}
S_{n+1}+I_{n+1}-(S_n+I_n)=-\gamma \delta_n I_n\,.\label{ngdSIR-C2}
\end{equation}
Combining (\ref{ngdSIR-C1}) and (\ref{ngdSIR-C2}), we obtain
\begin{eqnarray}
&&
\beta (S_{n+1}+I_{n+1}-(S_n+I_n))
-\gamma (\log S_{n+1}-\log S_n)\\
&& \qquad =-\beta \gamma \delta_n I_n
-\gamma \log\frac{1-(1-p_n)\beta \delta_n I_n}{1+p_n\beta \delta_n I_n}\,.\nonumber
\end{eqnarray}
If
\begin{equation}
\log\frac{1-(1-p_n)\beta \delta_n I_n}{1+p_n\beta \delta_n I_n}=-\beta \delta_n I_n
\end{equation}
is satisfied,
i.e., $p_n$ is given by (\ref{conditionformula2:ngdSIR}),
then (\ref{ngdSIR-CQ}) is a conserved quantity of the
nonautonomous gdSIR model (\ref{ngdSIR1}), (\ref{ngdSIR2}), (\ref{ngdSIR3}).
In other words,
\begin{eqnarray}
&& \frac{S_{n+1}-S_n}{\delta_n}
 =-\beta \left(\frac{(\beta \delta_nI_n -1)e^{\beta \delta_nI_n}+1}
{\beta \delta_nI_n(e^{\beta \delta_nI_n}-1)}(S_{n+1}+S_n)
+S_n\right)I_n\,,\label{ngdSIR1:int}\\
&& \frac{I_{n+1}-I_n}{\delta_n}
=\beta \left(\frac{(\beta \delta_nI_n -1)e^{\beta \delta_nI_n}+1}
{\beta \delta_nI_n(e^{\beta \delta_nI_n}-1)}(S_{n+1}+S_n)
+S_n\right) I_n  -\gamma I_{n}\,,\label{ngdSIR2:int}\\
&& \frac{R_{n+1}-R_n}{\delta_n}=\gamma I_{n}\,,\label{ngdSIR3:int}
\end{eqnarray}
has the same conservation quantities as the SIR model
(\ref{SIR1}), (\ref{SIR2}), (\ref{SIR3}),
thus we can think that the system of difference equations
(\ref{ngdSIR1:int}), (\ref{ngdSIR2:int}), (\ref{ngdSIR3:int})
with the hodograph transformation is an integrable discrete analogue of the SIR model.

\section{Integrability of an ultradiscretizable SIR model}

Sekiguchi et al. presented an ultradiscrete SIR model with time delay
and studies its analytical property~\cite{Sekiguchi}.
Although they presented some special solutions,
they did not mention about integrability of the ultradiscrete SIR model.
In this section, we consider an ultradiscretizable SIR model and its
ultradiscretization from the point of view of
integrability.

Let us  consider the following discrete SIR model:
\begin{eqnarray}
&& \frac{S_{n+1}-S_n}{\delta_n}=-\beta S_{n+1}I_n\,, \label{dSIR1:ud}\\
&& \frac{I_{n+1}-I_n}{\delta_n}=\beta S_{n+1} I_n -\gamma I_{n+1}\,, \label{dSIR2:ud}\\
&& \frac{R_{n+1}-R_n}{\delta_n}=\gamma I_{n+1}\,,\label{dSIR3:ud}
\end{eqnarray}
where  $S_n=S(\tau_n)$, $I_n=I(\tau_n)$, $R_n=R(\tau_n)$.
If $\delta_n$ is a constant,
this discrete SIR model is
a special case of the discrete SIR model considered in Sekiguchi et al.
This can be written as
\begin{eqnarray}
&& S_{n+1}=\frac{S_n}{1+\beta \delta_n I_n}\,,\label{dSIR1-2:ud}\\
&& I_{n+1}=\frac{(1+\beta \delta_n S_{n+1})I_n}{1+\gamma \delta_n}\,,
\label{dSIR2-2:ud}\\
&& R_{n+1}=R_n+\gamma \delta_n I_{n+1}\,.\label{dSIR3-2:ud}
\end{eqnarray}

Let us define $\tau_n$ and $t_n$ as (\ref{taudef}) and (\ref{tdef}).
Then we consider the discrete hodograph transformation
(\ref{dhodograph}) and the inverse discrete hodograph transformation (\ref{inversedhodograph}).
By using the relation $\epsilon_n=\delta_nI_n$,
the discrete SIR model (\ref{dSIR1:ud}), (\ref{dSIR2:ud}), (\ref{dSIR3:ud}) is transformed to
the following form:
\begin{eqnarray}
&& \frac{S_{n+1}-S_{n}}{\epsilon_n}=-\beta S_{n+1}\,, \label{udS}\\
&& \frac{I_{n+1}-I_n}{\epsilon_n}=\beta S_{n+1}  -\gamma \frac{I_{n+1}}{I_n}\,,\label{udI}\\
&& \frac{R_{n+1}-R_n}{\epsilon_n}=\gamma \frac{I_{n+1}}{I_n}\,.\label{udR}
\end{eqnarray}
We note that (\ref{udS}) is a linear difference equation
but (\ref{udI}) is a nonlinear difference equation.


Next we consider conserved quantities of the discrete SIR model
(\ref{dSIR1:ud}), (\ref{dSIR2:ud}), (\ref{dSIR3:ud}).
We can easily see that $S_n+I_n+R_n$ is a conserved quantity
of the discrete SIR model
(\ref{dSIR1:ud}), (\ref{dSIR2:ud}), (\ref{dSIR3:ud}).

By taking the logarithm of both sides of (\ref{dSIR2-2:ud}), we obtain
\begin{equation}
 \log S_{n+1}-\log S_n= -\log (1+\beta \delta_n I_n)\,.\label{udSIR-C1}
\end{equation}
Adding (\ref{dSIR1:ud}) and (\ref{dSIR2:ud}), we obtain
\begin{equation}
S_{n+1}+I_{n+1}-(S_n+I_n)=-\gamma \delta_n I_{n+1}\,.\label{udSIR-C2}
\end{equation}
Combining (\ref{udSIR-C1}) and (\ref{udSIR-C2}), we obtain
\begin{equation}
 -(S_{n+1}+I_{n+1}-(S_n+I_n))\log (1+\beta \delta_n I_n)
+\gamma \delta_n I_{n+1} (\log S_{n+1}-\log S_n)=0
\end{equation}
is zero for any $n$, this is invariant for
(\ref{dSIR1:ud}), (\ref{dSIR2:ud}), (\ref{dSIR3:ud}).

Setting $\epsilon_n=\delta_nI_n=\epsilon$ and $\delta_nI_{n+1}=\mu$, where
$\epsilon$ and $\mu$ are constant,
we obtain
\begin{equation}
 -(S_{n+1}+I_{n+1})\log (1+\beta \epsilon )
+\gamma \mu (\log S_{n+1})=-(S_{n}+I_{n})\log (1+\beta \epsilon )
+\gamma \mu (\log S_{n})\,,
\end{equation}
which indicates that
\begin{equation}
 (S_{n}+I_{n})\log (1+\beta \epsilon )
-\gamma \mu (\log S_{n})
\end{equation}
is a conserved quantity of the discrete SIR model
(\ref{dSIR1:ud}), (\ref{dSIR2:ud}), (\ref{dSIR3:ud}).
This means that the discrete SIR model (\ref{dSIR1:ud}), (\ref{dSIR2:ud}), (\ref{dSIR3:ud})
is integrable when $\epsilon_n$ and $I_{n+1}/I_{n}$ are constants.
Other cases including $\delta_n=\delta$, where $\delta$ is a constant, are nonintegrable.

In the case of $\epsilon_n=\epsilon$ and $I_{n+1}/I_{n}=\mu/\epsilon$, i.e., integrable case,
the discrete SIR model (\ref{dSIR1:ud}), (\ref{dSIR2:ud}), (\ref{dSIR3:ud}) is written as
\begin{eqnarray}
&& \frac{S_{n+1}-S_{n}}{\delta_n}=-\beta S_{n+1}I_n\,, \label{dSIR1:ud2}\\
&& \frac{I_{n+1}-I_n}{\delta_n}=\beta S_{n+1}I_n  -\gamma \frac{\mu}{\epsilon}I_n\,,\label{dSIR2:ud2}\\
&& \frac{R_{n+1}-R_n}{\delta_n}=\gamma \frac{\mu}{\epsilon}I_n\,,\label{dSIR3:ud2}
\end{eqnarray}
which is equivalent to the dSIR2 model.
Note that one of the conditions of integrability, $I_{n+1}/I_{n}=\mu/\epsilon$,
is too strong because this condition indicates that $I_n$ is a geometric sequence but this
is incompatible with the conservation of population.

Let us consider the solution to the initial value problem for the discrete
SIR model (\ref{dSIR1:ud}), (\ref{dSIR2:ud}), (\ref{dSIR3:ud}).
For the initial value
$S(t_0)=S(\tau_0)=S_0$, $I(t_0)=I(\tau_0)=I_0$, $R(t_0)=R(\tau_0)=R_0$,
the solution of (\ref{udS}), (\ref{udI}), (\ref{udR}) is given
by
\begin{eqnarray}
 && S_n=S_0\prod_{k=0}^{n-1}\frac{1}{1+\beta \epsilon_k}\,,\\
  &&I_n=I_0+
  \beta S_0\sum_{k=0}^{n-1}\epsilon_k
  \prod_{l=0}^{k}\frac{1}{1+\beta \epsilon_l}
  -\gamma \sum_{k=0}^{n-1}\epsilon_k\frac{I_{k+1}}{I_k}\\
  &&\quad =I_0+\beta \sum_{k=0}^{n-1} \epsilon_k S_{k+1}
  -\gamma \sum_{k=0}^{n-1}\epsilon_k\frac{I_{k+1}}{I_k}\,,\nonumber\\
  && R_n=R_0+\gamma \sum_{k=0}^{n-1}\epsilon_k\frac{I_{k+1}}{I_k}\,.
\end{eqnarray}
Substituting $\epsilon_n=\delta_n I_n$ into this solution,
the solution to the initial value problem for the
discrete SIR model
(\ref{dSIR1:ud}), (\ref{dSIR2:ud}), (\ref{dSIR3:ud})
is given by
\begin{eqnarray}
 && S_n=S_0\prod_{k=0}^{n-1}\frac{1}{1+\beta \delta_kI_k}\,,\\
  &&I_n=I_0+
    \beta S_0\sum_{k=0}^{n-1}\epsilon_k
    \prod_{l=0}^{k}\frac{1}{1+\beta \delta_{l}I_{l}}
    -\gamma \sum_{k=0}^{n-1}\delta_kI_{k+1}\label{dI-sol:ud}\\
  &&\quad =I_0+\beta \sum_{k=0}^{n-1}  \delta_kI_k S_{k+1}
    -\gamma \sum_{k=0}^{n-1}\delta_kI_{k+1}
\,,\nonumber\\
  && R_n=R_0+\gamma \sum_{k=0}^{n-1}\delta_kI_{k+1}\,\,,\\
 &&t_n=t_0+\sum_{k=0}^{n-1}\delta_k\,.
\end{eqnarray}
From (\ref{dI-sol:ud}), $I_n$ can be written as
\begin{eqnarray}
 I_n&=&\frac{\displaystyle I_0+\beta \sum_{k=0}^{n-1}\delta_k I_k S_{k+1}
-\gamma \sum_{k=0}^{n-2}\delta_kI_{k+1}}{1+\gamma \delta_{n-1}}\\
 &=&
\frac{\displaystyle I_0+\beta \sum_{k=0}^{n-1}\delta_k I_k S_0
\prod_{l=0}^{k}\frac{1}{1+\beta \delta_lI_l}-\gamma \sum_{k=0}^{n-2}\delta_kI_{k+1}}
{1+\gamma \delta_{n-1}}\,.\nonumber
\end{eqnarray}
From (\ref{dSIR2-2:ud}), $I_n$ can be also written as
\begin{equation}
I_n=I_0\prod_{k=0}^{n-1}\frac{1+\beta \delta_k S_{k+1}}{1+\gamma \delta_k}\,.
\end{equation}

If we set $\epsilon_n=\delta_nI_n=\epsilon$,
the above exact solution takes the following simpler form:
\begin{align}
 S_n&=\frac{S_0}{(1+\beta \epsilon)^n}\,,\\
I_n &=
 \frac{\displaystyle I_0+\beta \epsilon\sum_{k=0}^{n-1}
 \frac{S_0}{(1+\beta \epsilon)^{k+1}}-\gamma \sum_{k=0}^{n-2}\epsilon \frac{I_{k+1}}{I_k}}
 {1+\gamma \delta_{n-1}}\\
&= \frac{\displaystyle S_0+
I_0-S_0(1+\beta \epsilon)^{-n}
  -\gamma \sum_{k=0}^{n-2}\epsilon \frac{I_{k+1}}{I_k}}
  {1+\gamma \delta_{n-1}}
  =I_0\prod_{k=0}^{n-1}\frac{\displaystyle I_k+\beta \epsilon \delta_k \frac{S_0}{(1+\beta \epsilon)^{k+1}}}
{I_k+\gamma \epsilon}  \,,\notag\\
R_n&=R_0+\gamma \sum_{k=0}^{n-1}\epsilon \frac{I_{k+1}}{I_k}\,,\\
t_n&=t_0+\sum_{k=0}^{n-1}\delta_k\,.
\end{align}
Note that $S_n$ is written in the form of a power function which includes the infection rate $\beta$,
the lattice parameter $\epsilon$ and
the initial value $S_0$, but $I_n$ is not simple as the previous discrete SIR models,
i.e., for getting a solution, we need to compute the above formula recursively. This is due to
the lack of the second conserved quantity.

If we set $\delta_n=\delta$, the above solution leads to
\begin{align}
S_n&=S_0\prod_{k=0}^{n-1}\frac{1}{1+\beta \delta I_{k}}\,,\\
I_n&=I_0+\beta \delta \sum_{k=0}^{n-1} I_k S_{k+1}
-\gamma \delta \sum_{k=0}^{n-1}I_k\\
&=I_0+\beta  \delta S_0
\sum_{k=0}^{n-1}I_k \prod_{l=0}^{k}\frac{1}{1+\beta \delta I_{l}}
-\gamma \delta \sum_{k=0}^{n-1}I_k
=I_0\prod_{k=0}^{n-1}\frac{\displaystyle 1+\beta \delta S_0 \prod_{l=0}^{k}\frac{1}{1+\beta \delta I_{l}}}{1+\gamma \delta}\,,\notag \\
R_n&=R_0+\gamma \delta \sum_{k=0}^{n-1}I_{k+1}\,.
\end{align}

Setting $S_n=\exp(\mathcal{S}_n/h)$, $I_n=\exp(\mathcal{I}_n/h)$, $R_n=\exp(\mathcal{R}_n/h)$,
$\beta=\exp(\mathcal{B}/h)$,
$\gamma=\exp(\Gamma/h)$, $\epsilon_n=\exp(\mathcal{E}_n/h)$, $\delta_n=1$,
we obtain
\begin{eqnarray}
&& \exp(\mathcal{S}_{n+1}/h)=\frac{\exp(\mathcal{S}_n/h)}
{1+\exp((\mathcal{B}+\mathcal{I}_n)/h)}\,,\\
&& \exp(\mathcal{I}_{n+1}/h)
=\frac{
(1+\exp((\mathcal{B}+\mathcal{S}_{n+1})/h))
\exp(\mathcal{I}_n/h)
}
{1+\exp(\Gamma/h)}
\,,\\
&& \exp(\mathcal{R}_{n+1}/h)=\exp(\mathcal{R}_n/h)+\exp((\Gamma+\mathcal{I}_{n+1})/h)\,.
\end{eqnarray}
Taking the logarithm of both sides, we obtain
\begin{eqnarray}
&& \mathcal{S}_{n+1}=\mathcal{S}_n-h\log(1+\exp(( \mathcal{B}+\mathcal{I}_n)/h))\,,\\
&& \mathcal{I}_{n+1}= \mathcal{I}_n +
h\log(1+\exp((\mathcal{B}+\mathcal{S}_{n+1})/h))
-h\log(1+\exp(\Gamma/h))\,,\\
&&  \mathcal{R}_{n+1}=\mathcal{R}_n+\exp((\Gamma+ \mathcal{I}_{n+1})/h))\,.
\end{eqnarray}
Taking the ultradiscrete limit $h\to +0$,
we obtain the ultradiscrete SIR model
\begin{eqnarray}
&&  \mathcal{S}_{n+1}=\mathcal{S}_n-\max(0,\mathcal{B}+\mathcal{I}_n)\,,\\
&&\mathcal{I}_{n+1}= \mathcal{I}_n+
\max(0,\mathcal{B}+\mathcal{S}_{n+1})-\max(0,\Gamma)\,,\\
&& \mathcal{R}_{n+1}=\max(\mathcal{R}_n,\Gamma+\mathcal{I}_{n+1})\,
\end{eqnarray}
which is a special case of the ultradiscrete SIR model with time-delay~\cite{Sekiguchi}.
Here we used the following formula~\cite{ultra}:
\begin{equation}
 \lim_{h\to +0}\log\left(\exp\frac{A}{h}+\exp\frac{B}{h}\right)=\max(A,B)\,, \quad A, B>0\,.
\end{equation}
Note that the above ultradiscrete SIR model is nonintegrable because the discrete SIR model
(\ref{dSIR1:ud}), (\ref{dSIR2:ud}), (\ref{dSIR3:ud}) does not
have the second conserved quantity, but we can think that this is very close to
an integrable system.

\section{Conclusions}

We have presented structure-preserving discretizations of the SIR model,
namely the dSIR1, dSIR2, gdSIR and nonautonomous gdSIR models,
and their conserved quantities and exact solution to the initial value problem.
For these discretizations of the SIR model, the conditions for integrability have
been presented.
By choosing the best value of the parameter $p$ (for the gdSIR model) or
$p_n$ (for
the nonautonomous gdSIR model), the gdSIR model and
the nonautonomous gdSIR mode conserve
the conserved quantities of the continuous SIR model.
This fact suggests that the gdSIR model and the nonautonomous gdSIR model are
very powerful when used numerically.

We have also investigated an ultradiscretizable discrete SIR model and
its ultradiscretization, and
we conclude that the ultradiscretizable SIR model and its ultradiscretization
are not integrable.
However, it has some good properties that might make it near-integrable.

To the best of our knowledge,
structure-preserving discretizations of the SIR model
focusing on hodograph transformations have been previously unknown.
Our results may shed new light on the study of structure-preserving discretization
of mathematical models such as infectious diseases.
It is very interesting to extend our discretization method to
construct structure-preserving discretizations of other epidemic models such as the
SEIR model.

One of the authors (K.M.) would like to thank the Isaac Newton Institute for Mathematical Sciences, Cambridge, for support and hospitality during the programme "Dispersive hydrodynamics: mathematics, simulation and experiments",
with applications in nonlinear waves where work on this paper was undertaken.
This work was partially supported by JSPS KAKENHI Grant Numbers 18K03435,
22K03441, JST/CREST, and EPSRC grant no EP/R014604/1.
This work was also supported by the Research Institute for Mathematical Sciences,
an International Joint Usage/Research Center located in Kyoto University.


\end{document}